\documentclass[fleqn,usenatbib]{mnras}

\usepackage{newtxtext,newtxmath}

\usepackage[T1]{fontenc}

\DeclareRobustCommand{\VAN}[3]{#2}
\let\VANthebibliography\thebibliography
\def\thebibliography{\DeclareRobustCommand{\VAN}[3]{##3}\VANthebibliography}

\usepackage{graphicx}	
\usepackage{amsmath}	
\usepackage{xcolor,soul}

\newcommand{\bascs}{{\tt BASCS}}
\newcommand{\ebascs}{{\tt eBASCS}}
\newcommand{\spatial}{{\tt spatial}}
\newcommand{\spacetime}{{\tt space+time}}

\newcommand{\chandra}{{\it Chandra}}
\newcommand{\uvcet}{UV\,Cet}
\newcommand{\hbc}{HBC\,515}

\title[eBASCS]{eBASCS: 
Disentangling Overlapping Astronomical Sources II, using Spatial, Spectral, and Temporal Information}
\author[A. D. Meyer et al.]{Antoine D. Meyer,$^{1}$\thanks{E-mail: adm18@ic.ac.uk}
David A. van Dyk,$^{1}$
Vinay L. Kashyap,$^{2}$
Luis F. Campos,$^{3}$
David E. Jones,$^{4}$
\newauthor
Aneta Siemiginowska,$^{2}$
Andreas Zezas$^{2,5}$
\\
$^{1}$Imperial College London, Statistics Section, Department of Mathematics, 180 Queen’s Gate, London, UK SW7 2AZ\\
$^{2}$Center for Astrophysics $|$ Harvard \& Smithsonian, 60 Garden Street, Cambridge, MA 02138\\
$^{3}$Harvard University, Department of Statistics, Cambridge, MA 02138\\
$^{4}$Texas A\&M University, Department of Statistics, College Station, TX 77843\\
$^{5}$Department of Physics, University of Crete, 710 03 Heraklion, Crete, Greece
}

\date{Accepted 2021 May 17. Received 2021 May 17; in original form 2020 December 30}

\pubyear{2020}

\begin{document}
\label{firstpage}
\pagerange{\pageref{firstpage}--\pageref{lastpage}}
\maketitle

\begin{abstract}
The analysis of individual X-ray sources that appear in a crowded field can easily be compromised by the misallocation of recorded events to their originating sources. Even with a small number of sources, that nonetheless have overlapping point spread functions, the allocation of events to sources is a complex task that is subject to uncertainty.
We develop a Bayesian method designed to sift high-energy photon events from multiple sources with overlapping point spread functions, leveraging the differences in their spatial, spectral, and temporal signatures.  The method probabilistically assigns each event to a given source. Such a disentanglement allows more detailed spectral or temporal analysis to focus on the individual component in isolation, free of contamination from other sources or the background.
We are also able to compute source parameters of interest like their locations, relative brightness, and background contamination, while accounting for the uncertainty in event assignments.
Simulation studies that include event arrival time information demonstrate that the temporal component improves event disambiguation beyond using only spatial and spectral information.
The proposed methods correctly allocate up to 65$\%$ more events than 
the corresponding algorithms that ignore event arrival time information.
We apply our methods to two stellar X-ray binaries, \uvcet\ and \hbc\,A, observed with \chandra. We demonstrate that our methods are capable of removing the contamination due to a strong flare on \uvcet\,B in its companion $\approx$40$\times$ weaker during that event, and that evidence for spectral variability at timescales of a few ks can be determined in \hbc\,Aa and \hbc\,Ab.
\end{abstract}




\section{Introduction}
Analysis of X-ray data relies on the identification of the emitting sources and the allocation of the recorded events to the separated sources. When observing clusters of multiple contiguous sources, the presumable overlap of the sources' point-spread functions (PSFs) casts uncertainty on the true origin of each of the recorded events, as well as on the physical and spectral properties of components in the observed system. These uncertainties are often amplified by issues such as low-count data and
background contamination.
One possible solution for analysis involves fitting multiple PSF components to binned images \citep{2014ApJ...796...24P}. 
However, X-ray data are originally acquired as event lists that allow high spatial resolution, and also contain energy and arrival time information that is lost when binned into a 2D image. 
The purpose of the separated extraction regions is to allow further processing and individual analysis of the separated sources. This is sub-optimal in the event of substantial overlap in the sources' PSF, since events from each source are highly likely to contaminate the core of the other sources, which is expected to result in misclassification. For a given PSF size, this misclassification rate increases with both the size of the extraction region and the proximity of the sources. Moreover, events outside of the extraction regions are discarded from the analysis, which further inflates the uncertainties in parameter estimation due to the misclassification.

The problem with identification of sources in close pairs or in crowded fields is often encountered when running standard detection algorithms
\citep[e.g., wavdetect,][]{2002ApJS..138..185F} 
leading to source confusion and misclassification of extended sources. Source catalogs often flag problematic sources, use optical catalogs as a reference, or modify source 
and background extraction regions
to exclude the overlapping part \citep{Watson2009,Evans2010,2017A&A...598A...8P}.

The limitations of the manual extraction approach have motivated the development of alternative algorithmic methodologies to tackle the problem of overlapping point sources in high-energy photon image analysis. 
\cite{2015ApJ...808..137J} developed a statistical approach to probabilistically allocate events to sources using spatial and spectral information. This method is known as Bayesian Separation of Close Sources (\bascs{}). \bascs{} models the spatial distribution using a known PSF and leverages the differences in the energy spectra of the components to estimate the locations of the point sources and their relative intensities.
\bascs{} simultaneously provides the posterior distribution of the number of sources, which is particularly useful for detecting sources. This final feature of \bascs{} has added to the effort of probabilistic cataloging in dense fields that has been pursued by others (such as \cite{2017AJ....154..132P}, whose method applies to multi-band optical data), but is not the focus of this article. 
More recently, 
\citet{sottosanti2017discovering} used
a Bayesian mixture model to disentangle sources using spatial information, with the added feature of allowing for a diffuse non-isotropic background which is more suitable for $\gamma$-ray data. \citet{Picquenot_2019} developed a method to separate extended sources 
from the background by using spectral information (unlike \bascs{}, this method uses binned images and assumes that the source components all have similar spectra).
\citet{2019ApJ...877...17F} developed BayMAX, a statistical tool that uses spatial and spectral information to distinguish between single and dual active galactic nuclei (AGNs) via a Bayes Factor evaluation.

Here, we tackle the separate problem of time-variable sources.
In many cases, temporal information carries significant discriminatory power, since astronomical objects display brightness and spectral variability. Their spectral energy distributions are also expected to change with source intensity. Therefore, we expect a method that leverages the temporal signatures of the observed sources to outperform existing algorithms in the task of allocating recorded events to 
temporally variable sources. The principles behind the methodology developed here follow those of \cite{2015ApJ...808..137J}, but the variability of source intensity across time is incorporated into the model to aid in the task of source separation. 

The paper is organized as follows. Section~\ref{sec:data_and_statistical_models} provides an overview of the existing methodology and introduces the proposed methods.
Section~\ref{sec:inferenceandcomputation} presents the
statistical analysis framework and computational procedures.
Simulation studies are carried out in Section~\ref{sec:simulation_studies} to evaluate
the performance of the methods.
In Sections~\ref{sec:uvcet} and \ref{sec:hbc515} two datasets from observations of the \uvcet{} and \hbc\,A systems (respectively) are analyzed. Finally, potential extensions and limitations are discussed in Section~\ref{sec:conc} and detailed results from our numerical studies appear in a number of appendices.

\begin{table*}
\centering
\small
\begin{tabular}{ |l|l| }
    \hline
    Symbol & Definition\\
    \hline
    $(x_i, y_i)$ & Location of event $i$ on detector\\
    $E_i$ & Energy of event $i$\\
    $t_i$ & Event time of event $i$\\
    $l_i$ & Estimated wavelength of event $i$ \\
    $u_j = \{u_{j0}, ..., u_{jK}\}$ & A collection of pre-selected $K+1$ breakpoints defining time bins for source $j$\\
    $b_{ij}$ & An indicator of time bin for event $i$ and source $j$,\\
    & \quad\quad $b_{ij} = k$ if $t_i \in (u_{j,k-1}, u_{jk}]$\\
    $\mu_j$ & True {\emph{unknown}} location of source $j$ (2-D coordinates)\\
    $f_{\mu_j}$ & point spread function centered at $\mu_j$\\
    $\theta_{jk}$ & Spectral parameters for source $j$ and time bin $k$\\
    $\theta_{j}$ & Spectral parameters for source $j$ in \bascs{} algorithm\\
    $E_{min}, E_{max}$ & Minimum and Maximum detected energy\\
    $\pi_j$ & Relative intensity of source $j$ ($j = 0$ for background)\\
    $S$ & Number of Sources assumed\\
    $s_i$ & The true source of event $i$\\
    $\lambda_{jk}$ & Relative source intensity for source $j$ and time bin $k$\\
    $n_{jk}$ & True number of events detected for source $j$ and time bin $k$\\
    $n_{j\cdot}$ & True number of events detected for source $j$, i.e., $n_{j\cdot} = \sum_{k = 1}^K n_{jk}$\\
    $n_{\cdot k}$ & Number of events detected for time bin $k$, i.e., $n_{\cdot k} = \sum_{j = 1}^S n_{jk}$. These are \\
    $n$ & Total number of events detected, i.e., $n_{\cdot k} = \sum_{j,k} n_{jk}$\\
    $(x, y, E, t, b)$ & Vectors of the corresponding event specific variables\\
    \hline
\end{tabular}
\caption{Symbols used in this work.}
\label{tab:my_label}
\end{table*}

\section{Data and Statistical Models} 
\label{sec:data_and_statistical_models}
\subsection{Structure of the Data}
\label{sec:structuredata}

High-energy detectors, such as Charged Coupled Device (CCD) imaging spectrometers, record detector spatial coordinates, $(x_i, y_i)$, arrival time, $t_i$, and energy, $E_i$. The full event list for $n$ recorded events is denoted as $\mathbf{x} = \{\mathbf{x}_i\}_{i=1}^n=\{(x_i,y_i,t_i,E_i)\}_{i=1}^{n}$. The observed spatial and spectral data is subject to the effect of the Point Spread Function (PSF) and the Redistribution Matrix Function (RMF). While the spatial dispersion of events by the PSF is explicitly accounted for in the model, for this study the RMF-induced energy dispersion is ignored and the observed spectra is modelled, rather than the source spectra.

Each recorded event is assumed to originate either from one of the sources located in the field of view,
or from the background. In this article, only point sources are considered. Our proposed methods, unlike \bascs{} \citep{2015ApJ...808..137J}, assume that the number of observed sources on the image is known. The location, intensities, spectral distributions, and light curves are unknown. 
Background events are assumed to be uniformly distributed 
spatially under the source.
Their spectrum is
assumed to be known up to a normalization factor, and can be either uniform, or one of the models described in Section~\ref{sec:spectralmodel}. 
We expect the 
modeling
to be robust to mild deviations from the
assumption of spatial uniformity of the background,
as photons that are spatially distant from the sources are highly likely to be attributed to the background, irrespective of the background model.

\subsection{Finite Mixture Model}\label{sec:FiniteMix}

A natural way to describe data assumed to originate from a collection of sub-populations (in this case, the sources and background) is with the class of statistical models known as \emph{finite mixture distributions}. In such models, each event $\mathbf{x}_i$, $i = 1,\dots,n$ is assumed to arise from one of $S+1$ ($S$ sources and the background) component distributions $\{h_j(\mathbf{x}_i|\Theta_k)\}_{j=0}^S$, each parametrized by a parameter vector $\Theta_k$. It is further assumed that it is not known which mixture component generated each observation. The background contamination is defined as corresponding to component $j=0$ of the finite mixture. Further denote by $w_j$ the proportion of the population originated from mixture component $j$, so that $\sum_{j=0}^{S} w_j=1$. The likelihood of data $\mathbf{x}$ is thus:
\begin{equation}
    p(\mathbf{x}_i|\Theta, \mathbf{w}) = \sum_{j=0}^S w_j h_j(\mathbf{x}_i|\Theta_j).
\end{equation}

With the aim of learning which mixture component underlies each observed event, we introduce the latent indicator variables, $s_i$, into the model: $s_i = j$ if $\mathbf{x}_i$ is drawn from mixture distribution $h_j$ i.e., if event $i$ originated from source $j$.
Since $\mathbf{w} = (w_0, \dots, w_S)$ gives the proportion of the population of events belonging to each mixture, and $\mathbf{s} = (s_{1}, \dots, s_{n})$ gives the mixture assignment of event $\mathbf{x}_i$, it is sensible to model $\mathbf{s}$ with a Multinomial \footnote{Let $n_j = \sum_{i=1}^n 1(s_i = j)$ for $j = 0, \dots, S$. The probability mass function of the Multinomial$(n;(w_0, \dots, w_S))$ distribution is then given by: $p(n_{0},\dots,n_{S}) = n / n_{0}!\dots n_{S}! \prod w_{0}^{n_{0}}\dots w_S^{n_{S}}$.} distribution:
\begin{equation}
    (s_1,\dots,s_n)\mid \mathbf{w} \sim \text{Multinomial}(n;(w_0,\dots,w_S)).
    \label{eq:BASCS_weights}
\end{equation}
The joint distribution of the data, $\mathbf{x},$ and latent variable, $\mathbf{s}$, is then given by:
\begin{align}
    p(\mathbf{x},s|\Theta, \mathbf{w}) &= p(\mathbf{s}|\mathbf{w})p(\mathbf{x}|\mathbf{s},\Theta) 
                           = \prod_{i=1}^n p(s_i|\mathbf{w})p(\mathbf{x}|s_i,\Theta) \nonumber\\
                           &= \prod_{i=1}^n \prod_{j=0}^S p(s_i = j|w_j)p(\mathbf{x}_i|s_i = j,\Theta_j) \nonumber\\
                           &= \prod_{i=1}^n \prod_{j=0}^S w_j h_j(\mathbf{x}_i|\Theta_j).
\end{align}
For the rest of this article, we refer to the mixture weights, $\mathbf{w}$, as the relative intensities of the sources. Now that the general framework for statistical modelling of overlapping sources is established, models for the individual mixture distributions $\{h_j(\mathbf{x}_i|\Theta_j)\}_{j=0}^S$ can be constructed. In particular, each component is a combination of sub-models for the spatial, spectral and temporal data. \bascs{} \citep{2015ApJ...808..137J} only included $\{(x_i, y_i, E_i)\}_{i=1}^n$, i.e., only the spatial and spectral information. The contribution to the existing framework is the formulation of a model for the temporal data $\{t_i\}_{i=1}^n$ and its integration into the current methodology.

\subsection{Spatial Model}

The detector spatial coordinates of an event $(x_i, y_i)$ are a deviation from the actual (unknown) position of the event's originating source (say $j$), $\mu_j = (\mu_{xj}, \mu_{yj})$.
This deviation is due to the PSF, the response of the imaging system to the incident events that redistributes them on the surface of the detector according to a telescope and detector specific distribution centered at $\mu_j$.
In our numerical studies, 
we assume that the PSF is well approximated by a King profile 
\citep[see Appendix~\ref{appsec:king} and ][]{2015ApJ...808..137J}.
The observed event locations from source $j$ are, therefore, distributed according to the PSF centered at the unknown source location $\mu_j$, i.e.,
\begin{align}
\label{eq:BASCS_spatial}
    (x_i, y_i) \mid (s_i = j, \mu_j) \sim f_{\mu_j}(x_i,y_i)
\end{align}
for $j = 1,\dots,S$, where $f_{\mu_j}(x_i,y_i)$ denotes the PSF centered at $\mu_j$ evaluated at $(x_i, y_i)$. 

As for the background, the assumption of spatial uniformity across the image means:
\begin{equation}
    (x_i, y_i)\mid(s_i = 0) \sim \text{Uniform.}
\end{equation}

\subsection{Spectral Model}
\label{sec:spectralmodel}
In addition to spatial information, \cite{2015ApJ...808..137J} discuss the benefits of including spectral information when allocating events to sources. Generally, approximating the observed spectral energy distribution\footnote{
Since we use 
approximate
spectral shapes to distinguish sources, we ignore 
the effects of the Redistribution Matrix Functions (RMF) and Auxiliary Response File (ARF) and instead model the distributions of the observed 
spectrum as recorded by the detector. 
Although differences in the observed spectra are just as powerful as differences in source spectra for source separation, 
it precludes us from interpreting
the parameters of the observed spectral model in terms of the physical properties of the sources.
}
 with a rough shape is a simple way to exploit spectral differences among sources, with relatively little additional computation. A simple sensible photon energy model is a Gamma distribution\footnote{The density of the Gamma$(\alpha, \alpha/\gamma)$ distribution is $f(x) = \frac{\alpha^{\alpha}}{\gamma^{\alpha}\Gamma(\alpha)}x^{\alpha-1} e^{-\frac{\alpha}{\gamma}x}$, $x > 0$.}, 
because it only allows positive energy values and its
mode is around the mid-to-low energy region. This yields:
\begin{equation}
    \label{eq:singlegamma}
    E_i \mid (s_i = j, \alpha, \gamma) \sim \text{Gamma}(\alpha_j, \alpha_j/\gamma_j)
\end{equation}
for $j = 1,\dots,S$,  where $\alpha_j$ and $\gamma_j$ are respectively the unknown shape and mean parameters of the Gamma distribution. This modeling strategy makes no attempt at describing the energy distributions in a detailed science-based manner, since for instance emission lines are ignored. Nevertheless, \cite{2015ApJ...808..137J} showed that when sources are relatively close spatially, modelling their rough spectral shapes improves source detection and separation and increases the precision of source parameter estimates.

In practice, however, a single Gamma distribution may not capture the shape of the spectra sufficiently well, and the miss-specification of the spectral model can bias the parameter estimates. \citet{2015ApJ...808..137J} discuss a more general spectral model, defined by a mixture of two Gamma distributions, that allows for a more flexible approximation of the observed energy distribution. This energy model can be written as:
\begin{align}
    \label{eq:mixgamma}
     E_i &\mid (s_i = j, \alpha_1, \alpha_2, \gamma_1, \gamma_2, \pi) \nonumber \\&\sim \pi_{j} \text{Gamma}(\alpha_{j1}, \alpha_{j1}/\gamma_{j1})+ (1- \pi_j) \text{Gamma}(\alpha_{j2}, \alpha_{j2}/\gamma_{j2})
\end{align}
for $j = 1,\dots,S$, where $\pi_j$ is the mixture weight. We denote the modelled spectral distribution by $g_\theta$, with $\theta$ representing its parameters. The observed spectral energy for source $j$ is therefore distributed as $g_{\theta_j}$ with unknown parameter $\theta_j$, i.e,
\begin{align}
\label{eq:BASCS_energy}
   E_i\mid(s_i = j, \theta) \sim g_{\theta_{j}}(E_i).
\end{align}
The reader is referred to \cite{2015ApJ...808..137J} for further discussion on the specification of the spectral model and the costs and benefits of more complex spectral energy distributions.

The background contamination is again assumed to have a uniform spectral distribution:
\begin{equation}
    E_i \mid s_i = 0 \sim \text{Uniform}(E_{min}, E_{max})
\end{equation}
where $E_{min}$ and $E_{max}$ are the detector-specific minimum and maximum observable photon energies.



\subsection{Temporal Model and Time-varying Energy Distributions}

Our proposed extension of \bascs{} \citep{2015ApJ...808..137J} incorporates event arrival times 
into the model. The assumption of \bascs{} that both the relative brightness of the sources and their spectral energy distributions remain invariant during the observation period is thus relaxed. 
In general, astronomical objects 
display brightness and spectral variability.
Furthermore, spectral energy distributions are also expected 
to change with source intensity. 
In many 
cases temporal information therefore carries significant discriminative power and including it will outperform \bascs{} in the allocation of recorded events to sources.

We model the light curve of each source as a piece-wise constant function defined on a pre-selected collection of $K$ time bins,  the union of which equals the entire observation period. Assuming that the light curves are uniform over each of the $K$ time intervals yields a flexible temporal model, suitable to flexibly capture 
temporal variability. 

We formulate the proposed extension as follows: for each source $j$, let $u = \{u_{k}\}_{k=0}^K$, with $0 = u_{0} < u_{1} < ... < u_{K} = T$, be a collection of breakpoints segmenting the full observation interval into $K$ bins. Let $b_i$ indicate the bin in which event $i$ is detected, i.e.,  
$b_{i} = k$ if $t_i \in (u_{k-1}, u_{k}]$.  The labels $\{b_{i}\}$ are an observed, discretized version of the original time-arrival data. These play a role in measuring the time-varying intensity within each of the sources. 

To measure the within-source time relative intensity, first let $n_{jk}$ represent the number of events from source $j$ that are observed in time bin $k$. For each source $j$, let $\lambda_{j} = (\lambda_{j1}, \lambda_{j2}, ..., \lambda_{jK})$ be the relative intensities for source $j$ across the $K$ time bins, with $\sum_{k = 1}^K \lambda_{jk} = 1$. For source $j$, the number of events observed across the time bins are modeled as a Multinomial distribution, 
\begin{align}
	\label{eq:eBASCS_time}
    (n_{j1}, n_{j2}, ..., n_{jK})\mid\lambda_{j}, n_j, S \sim \text{Multinomial}(n_j;\lambda_{j}).
\end{align}
where $\sum_{k = 1}^K n_{jk} = n_j$. The collection of time intensities
across the background and the $S$ sources
is denoted  $\lambda = (
\lambda_0
,\lambda_{1}, ..., \lambda_{S})$.

The temporal model is not intended to capture nuanced variability in the observed light curves. Instead, in conjunction with coarse energy distributions, it captures general trends in time that may have discriminatory power.

The source-level energy distributions may also 
be time variable, i.e., they may change at each of the breakpoints $u$. The energy distribution for events arriving in time bin $k$ from source $j$ is modeled with a distribution $g$ and source-time specific parameters $\theta_{jk}$, i.e.,
\begin{align}
\label{eq:eBASCS_energy}
    E_i\mid (s_i = j, b_{ij} = k) \sim g_{\theta_{jk}}(E_i),
\end{align}
where $g_\theta$ denotes the mixture of two Gamma distributions defined in Section~\ref{sec:spectralmodel}. 
In this case the models for the spectral energy distribution are independent for each source and within each time bin. Because the number of parameters in this energy model grows rapidly with the number of sources and the number of time bins, it is difficult to fit without a sufficient number of events. A more sophisticated model might have fewer time bins for dimmer sources. In our numerical studies and examples, we assume the source spectra are invariant over the whole observation period.

\subsection{Selection of time bins}
 \label{sub:binselection}

\begin{figure}
    \centering
    \includegraphics[width = \linewidth]{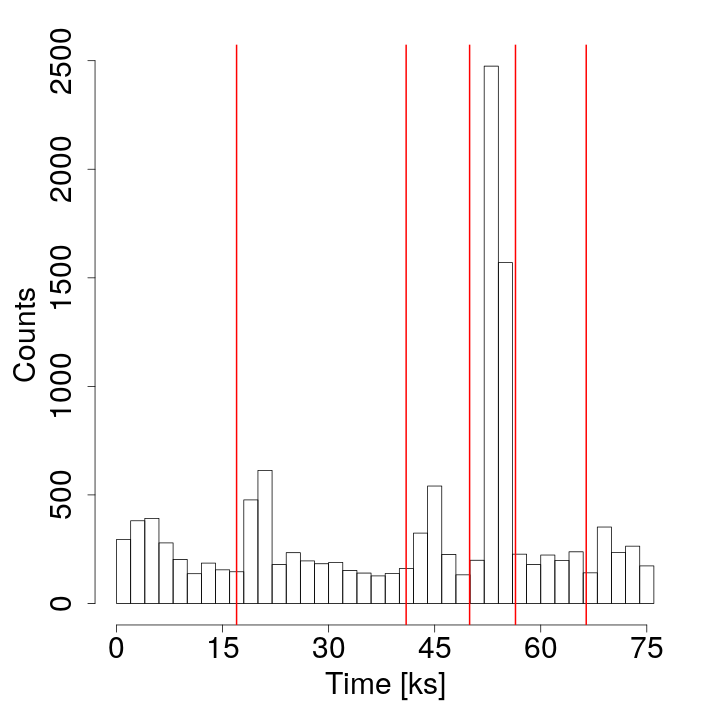}
    \caption{
    The light curve of the combined \uvcet\ system, with a time bin size of $\approx$ 1500~s. Our data-driven procedure is deployed to identify the five break points denoted by the red vertical bars. 
    }
    \label{fig:algoseg}
\end{figure}



In order to account for temporal changes, we split the data into possibly irregular bins that capture the variations in the observed light curve of the system.  There is a trade-off here between the fineness with which one wants the variability to be modeled on the one hand, and the data size that allows useful uncertainty intervals on the fitted parameters, the total number of parameters, and the running time. The dimension of the parameter space scales linearly with the number of time segments; for the sake of parsimony as well as to avoid unnecessarily complex calculations, a small number of segments is preferred.  The temporal segmentation model is not designed to capture subtle variations directly, but rather aims to improve disambiguation between overlapping photon events.

We thus allow the number of breakpoints to be set on a case by case basis, and choose as the breakpoints those times which show the largest changes in counts in adjacent bins. 
Given the user-specified number of breakpoints, we develop a simple
data-driven procedure to identify the locations that best
isolate transient variations and flares in the light curves.
The procedure first divides the event-arrival times into a high-resolution histogram and then identifies the adjacent bins with the largest differences in intensity. The breakpoints are set to separate these bins; see Figure~\ref{fig:algoseg} for an illustration of this method in the case of the \uvcet\ light curve with the number of breakpoints preset to five.
Finally, the algorithm merges all of the histogram bins between the selected breakpoints to obtain the target number of time intervals. Full details appear in the form of pseudo-code in Appendix~\ref{sec:apptimebin}; computer code is included in the \ebascs{} software package
available on the CHASC GitHub software library\footnote{\url{https://github.com/AstroStat/}}.



\subsection{Statistical Models and Likelihoods}

Here we combine the spatial, spectral, and temporal models into the overall models that we use and compare in this article.
We refer to the model that only incorporates spatial data, $\{(x_i, y_i)\}_{i=1}^n$, as the baseline or \spatial{} model. Its parameters are denoted by $\Theta = \{\mu_j\}_{j=1}^S$, and the corresponding likelihood is:
\begin{equation}
    L^{\spatial{}}(\Theta, s) \propto \prod_{i=1}^n \prod_{j=1}^S w_j f_{\mu_j}(x_i,y_i).
\end{equation}
 
\bascs{} \citep{2015ApJ...808..137J} models observed event spatial and energy data: $\{(x_i, y_i, E_i)\}_{i=1}^n$. The parameters for this model are denoted $\Theta = \{\mu_j, \theta_j\}_{j=1}^S$, and the corresponding likelihood is:
\begin{equation}
    L^{\bascs{}}(\Theta, s) \propto \prod_{i=1}^n \prod_{j=1}^S w_j f_{\mu_j}(x_i,y_i)g_{\theta_j}(E_i).
\end{equation}

The proposed method, \ebascs{}, models observed event spatial, spectral and temporal data: $\mathbf{x}=\{(x_i,y_i,t_i,E_i)\}_{i=1}^{n}$. The parameters for this model are denoted $\Theta = \{\mu_j, \theta_j, \lambda_j\}_{j=1}^S$, and the corresponding likelihood is:
\begin{equation}
    L^{\ebascs{}}(\Theta, s) \propto \prod_{i=1}^n \prod_{j=1}^S \prod_{k=1}^K w_j \lambda_{jk} f_{\mu_j}(x_i,y_i)g_{\theta_j}(E_i).
\end{equation}

Some X-ray detectors do not record the energy of the observed events 
with useful accuracy
, e.g., the \chandra/HRC-S detector used to record 
the \uvcet\ observation analysed in Section~\ref{sec:uvcet}. In such cases, it is useful to consider a spatio-temporal model 
that only includes the spatial and temporal components. We refer to this model as the \spacetime{} model. The parameter for this model is $\Theta = \{\mu_j, \lambda_j\}_{j=1}^S$, which yields the likelihood:
\begin{equation}
    L^{\spacetime{}}(\Theta, s) \propto \prod_{i=1}^n \prod_{j=1}^S \prod_{k=1}^K w_j \lambda_{jk} f_{\mu_j}(x_i,y_i).
\end{equation}


\section{Bayesian Inference and Statistical Computation}
\label{sec:inferenceandcomputation}

\subsection{Inferential approach}

We adopt a Bayesian statistical approach. This allows us to quantify uncertainty in the model parameters via their joint (posterior)
distribution given the observed data. From this we can compute point estimates and error bars, if required.  The Bayesian paradigm also allows us to incorporate existing scientific knowledge about the likely values of the parameters into the model via their prior distributions. Once the prior distributions are specified and the data is observed (and integrated into the model via the likelihood), Bayes' Theorem gives the expression for the posterior distribution of the parameters, i.e., the updated beliefs about the model parameters after observing the data:
\begin{equation}
    p(\Theta, s | \mathbf{x}) = \frac{p(\mathbf{x} | \Theta, s)p(\Theta, s)}{p(\mathbf{x})},
    \label{eq:bayesform}
\end{equation}
where $p(\mathbf{x} | \Theta, s) \equiv L(\Theta, s)$ is the likelihood of the data under the chosen model, $p(\Theta, s)$ is the prior distribution of the parameters, and $p(\mathbf{x})$ is the marginal distribution of the data.  While $p(\mathbf{x})$ is sometimes used for model selection, for the purposes of Equation~\ref{eq:bayesform}, it can be viewed as a normalizing constant and need not be computed. 
Our choices of prior distributions are discussed in Section~\ref{sec: priors}.


\subsection{Specifying prior distributions} \label{sec: priors}
Our general approach is to specify uninformative and computationally practical prior distributions. In our applications, we usually observe a large enough data set to mitigate the effect of the prior distributions on posterior inference. However, any information about likely parameter values can and should be encoded into the prior distributions. 

Following \citet{2015ApJ...808..137J}, we specify uniform (across the image) priors on the source locations: 
\begin{equation}
    \mu_j \sim \text{Uniform}
\end{equation}
for $j=1,\dots,S$ 
, but one could also specify a distribution centered at a likely value $\mu_{j0}$ for $\mu_j$.  The prior distributions for the other parameters are those given in \citet{2015ApJ...808..137J}. The parameters that are probability vectors, $\mathbf{w}$ and $\lambda$, and govern the Multinomial splitting of events among sources and time bins, respectively, are given Dirichlet prior distributions.
Picking conjugate prior distributions (like Dirichlet priors for Multinomial likelihoods) is computationally convenient, and allows for more transparent interpretations of posterior inferences \citep{gelman2013bayesian}. Hence:
\begin{align}
    &\mathbf{w} \sim \text{Dirichlet}(\eta, \dots, \eta), \nonumber\\
    &\lambda \sim \text{Dirichlet}(\zeta, \dots, \zeta).
\end{align}
Following \citet{2015ApJ...808..137J}, the above hyperparameters are set at 
$\eta  = \zeta = 1$, so that the prior distributions correspond to as much information as a single event added to each source (or each time bin for $\lambda$). 

For a spectral model defined as a single Gamma distribution, we use the prior distributions, $\alpha_j \sim \text{Gamma}(2,0.5)$ and $\gamma_j \sim \text{Uniform}(E_{min}, E_{max})$. If a mixture of two Gamma distributions is used for the spectral model, a reasonable choice for the mixture weight parameter is a  Beta(2,2) distribution (assuming a 2-components mixture), again following \citet{2015ApJ...808..137J}. 

\subsection{Statistical Computation}
\label{sec:comp}

In order to derive useful summaries of the posterior distribution of the model parameters, we generate a Monte Carlo sample of parameter values from the posterior. Indeed, for an unknown model parameter $\mu$ with marginal posterior distribution $p(\mu | \mathbf{x})$, a sample $\{\mu^{(1)}, \dots, \mu^{(N)} \}$ of $N$ draws from the posterior allows us to compute point estimates and error bars for $\mu$, e.g., via the mean and standard deviation of the sample 
if the posterior is roughly Gaussian
\citep[e.g.,][]{StenningDyk+2021+29+58}, or more generally via quantiles of the sample.  

We use Markov Chain Monte Carlo (MCMC) to obtain a sample of $(\Theta, s)$ from its joint posterior distribution, $p(\Theta, s | \mathbf{x})$. 
A Markov chain is a set of sequentially sampled random variables, $(\Theta,s)^{(t)}$, for $t=1,2,\ldots$, such that each $(\Theta,s)^{(t)}$, depends on the history of the chain only through the most recent iterate, $(\Theta,s)^{(t-1)}$.\footnote{More formally, $\{(\Theta,s)^{(1)}, \ldots, (\Theta,s)^{(t-1)}\}$ and $(\Theta,s)^{(t+1)}$ are conditionally independent given $(\Theta,s)^{(t)}$, for each $t$.}
MCMC methods are a class of iterative algorithms that produce a Markov chain with \emph{stationary distribution} equal to the target Bayesian posterior distribution. If run for a sufficient number of iterations, the marginal distribution of each of the (correlated) iterates of the chain approaches the target posterior distribution. Thus, if the sample that is comprised of the MCMC iterations before the chain reaches approximate convergence (i.e., the burn-in iterations) is discarded, the remaining sample (i.e., the main run) can be treated as a correlated sample from the target posterior distribution.
In our simulations, we typically discard the first half of the chain as burn-in.


Positively
correlated samples carry less information than independent samples of the same size, and thus yield estimates of posterior quantities with higher Monte Carlo error. The Effective Sample Size (ESS) approximates how large of an independent sample 
would contain the same level of information as our correlated MCMC sample  \citep[e.g.,][]{StenningDyk+2021+29+58}.  
In our numerical studies and examples, we choose the number of main-run iteration to obtain an ESS between 1000 and 2000. 
Generally, this requires a total run of about 40,000 iterations. (We discard the first half of each chain as burn-in and thin the remainder, saving only every tenth iterate, to reduce memory requirements.) We also monitor
convergence of the chains through visual diagnostics such as trace plots and auto-correlation plots.
For a more detailed discussion of practical considerations involved with MCMC,  the reader is referred to \citet{gelman2013bayesian} or, 
for a more astronomy-oriented account, to  \citet{StenningDyk+2021+29+58}. 
We develop both R code for MCMC methods based on the Metropolis algorithm\footnote{
The Metropolis algorithm is a simple MCMC sampler that produces a Markov chain by sampling the next iterate from a proposal distribution centred at the current iterate (i.e. a symmetric proposal distribution), using a rejection rule designed to ensure that the stationary distribution of the Markov chain equals the target posterior distribution.
See \citet{2019arXiv190912313S} for a recent, and accessible, description of MCMC concepts.
} \citep{1953JChPh..21.1087M} and STAN\footnote{
STAN is a probabilistic programming language for Bayesian inference with gradient-based MCMC techniques \citep{JSSv076i01}.
} code which implements Hamiltonian Monte Carlo\footnote{
Hamiltonian Monte Carlo is an MCMC method that exploits the differential structure of the target posterior distribution to generate a Markov chain that efficiently represents the posterior distribution. (See \cite{2017arXiv170102434B} for an introduction to Hamiltonian Monte Carlo.)
} (\citet{DUANE1987216}, \citet{doi:10.1201/b10905-7}). 

\begin{figure}
    \centering
    \includegraphics[width = 0.45\linewidth]{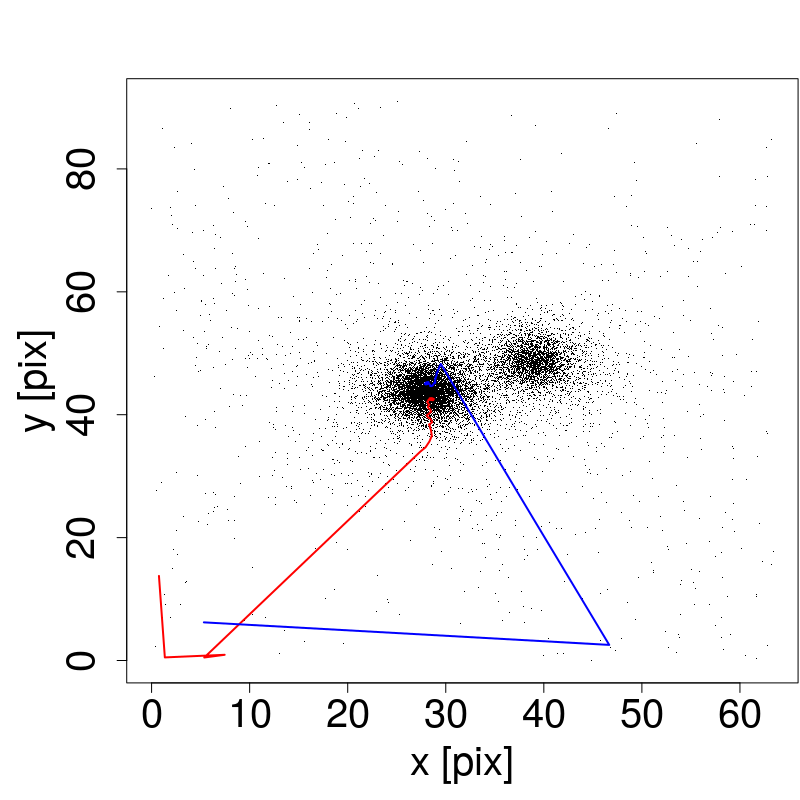}
    \includegraphics[width = 0.45\linewidth]{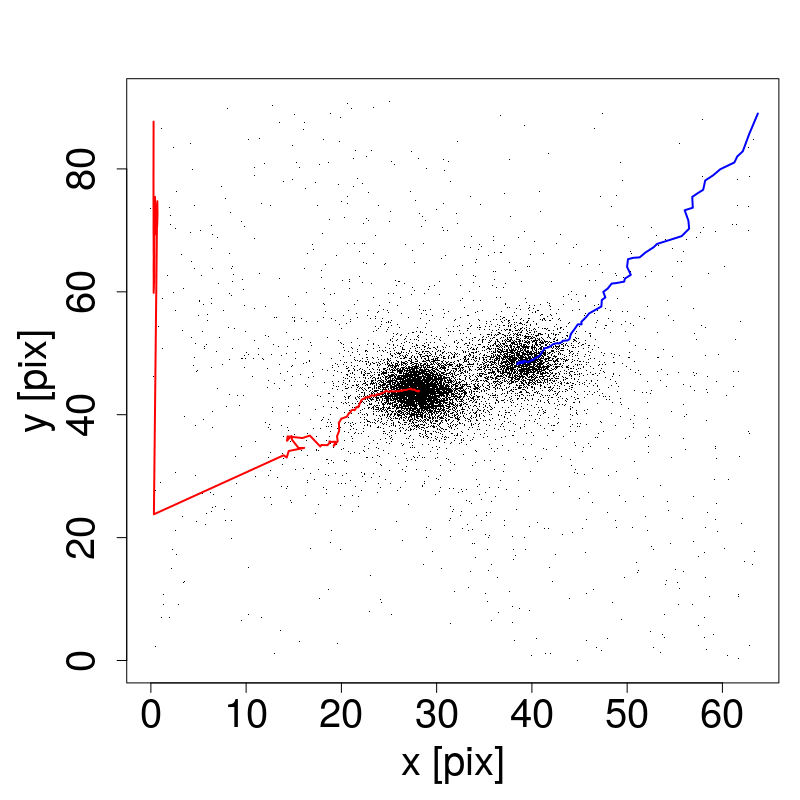}
    \caption{Examples of the convergence the location parameters' MCMC iterates. Both panels show the recorded spatial coordinates of the UV Ceti (Gliese 65) binary system, analyzed in Section \ref{sec:uvcet}.
    \emph{Left Panel:} With poor choice of initialization, both fitted location parameters, $\mu_1$ (red) and $\mu_2$ (blue), converge to the location of source to the left of the image.  \emph{Right Panel:} When better initialized, they converge to their respective true locations.}
    \label{fig:badmcmc}
\end{figure}

Naturally, our methodology has certain limitations. Here we detail two and propose solutions.
Finite mixture models, such as the one we use with \ebascs{}, produce multi-modal posterior distributions, which are notoriously difficult to explore via MCMC and thus  require special care. The most obvious and important issue is the possibility that the algorithm becomes "stuck" in one mode, and thus fails to explore the full posterior distribution. To give a more detailed example, consider a binary star system, with unknown location parameters $\mu_1$ and $\mu_2$. If the sources are well separated, it might happen that the MCMC samples of both $\mu_1$ and $\mu_2$ converge to the location of source 1, see Figure \ref{fig:badmcmc}. 
This worry can be 
addressed
by choosing 
starting values for $\mu_1$ and $\mu_2$ at random locations along the edges of the image close to the respective sources. Another option, that we recommend, is to initialize the location parameter values at the approximate location of the modes of the spatial image of observed event locations,
e.g., found by applying  Kernel Density Estimation\footnote{
Kernel Density Estimation is a non-parametric procedure designed to estimate probability densities from observed data \citep{Davis2011}. 
} (KDE) to the raw images. 
We have not found it necessary to implement sophisticated multi-modal sampling techniques (e.g., parallel tempering\footnote{
Parallel tempering is a multiple chain MCMC method, where each chain targets one of a sequence of tempered versions of the target distribution. Swapping draws among the chains enables improved exploration of multi-modal posterior distributions.
} \citep{geyer1991markov} or evolutionary Monte Carlo \footnote{
Evolutionary Monte Carlo incorporates features from simulated annealing and genetic algorithms to improve on the efficiency of standard MCMC methods.
} \citep{10.2307/24306722})
since the failure mode has been obvious and the remedies easily implemented.

Another possible concern is that
result in computational inefficiency.
For instance, the source locations $\mu_1$ and $\mu_2$  
can only take values within the boundaries of the image, and the spectral parameters must be positive. Using a standard Metropolis 
proposal distribution
for these parameters, such as a normal distribution centered on the current value, leads to severe inefficiency in the algorithm since proposed values outside of the parameter's support must be rejected, which induces sub-optimality in the Metropolis acceptance rate. To remedy this, we log-transform the parameters 
to eliminate the constraints on their support. This allows us to efficiently implement 
a normal 
proposal distribution
(on the logarithmic scale).




\begin{figure*}
    \centering
    \includegraphics[width = 0.32\linewidth]{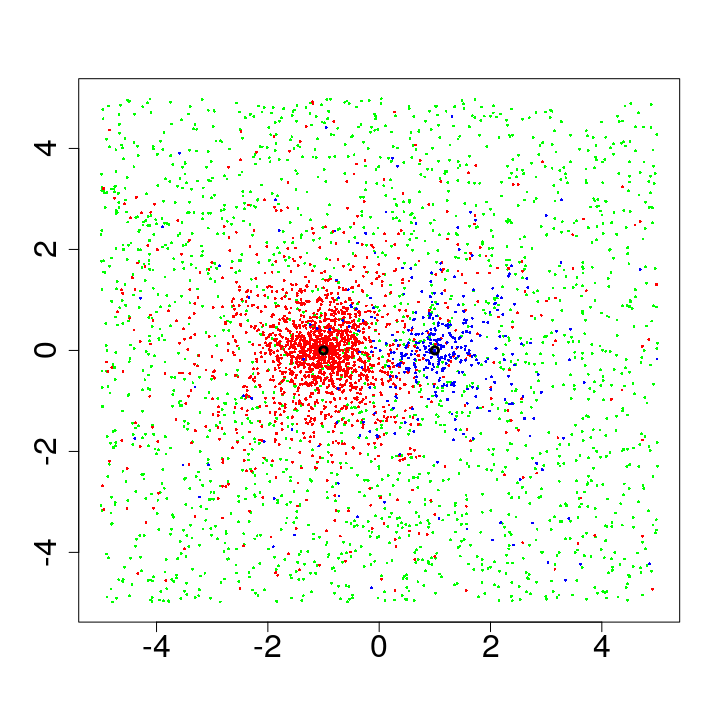}
    \includegraphics[width = 0.32\linewidth]{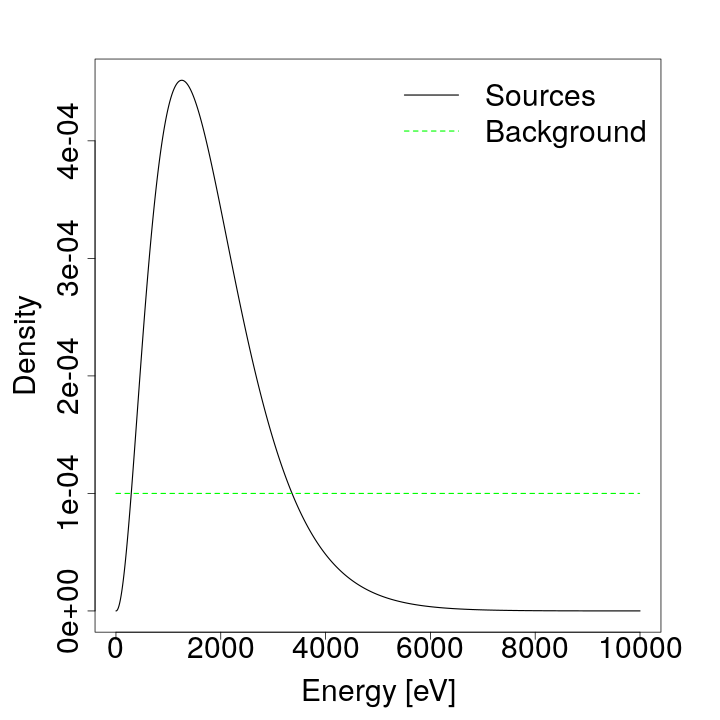}
    \includegraphics[width = 0.32\linewidth]{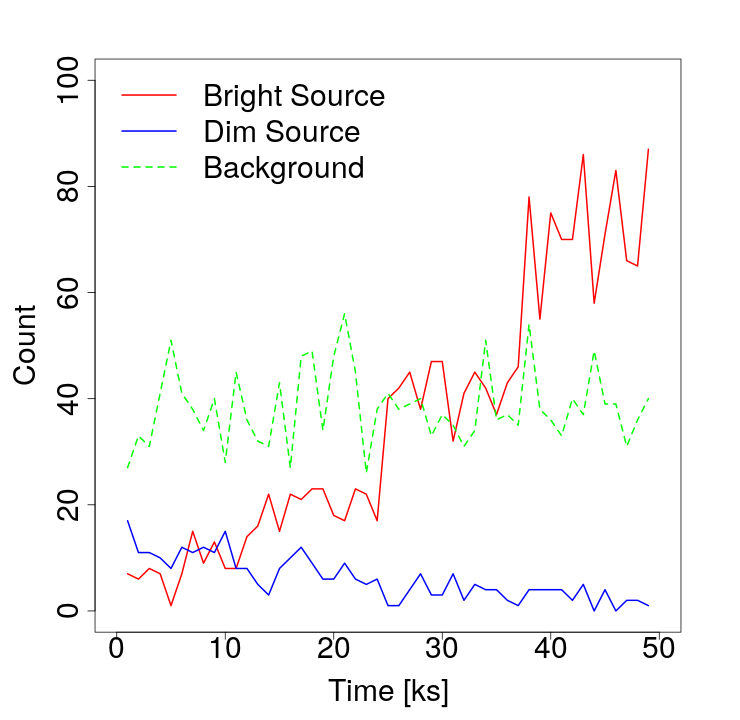}
    \caption{Illustrative simulation design, with parameters $r = 5$, $d=2$, $b=1$. \emph{Left Panel:} Scatter plot of simulated event locations for the bright source (red), the dim source (blue) and the background (green). True source locations are marked by the black circles. \emph{Middle Panel:} Spectra used to simulate the source and background event energies. \emph{Right Panel:} Observed light curves for the sources and background.}
    \label{fig:simsetup}
\end{figure*}

\begin{figure}
    \centering
    \includegraphics[width=\columnwidth]{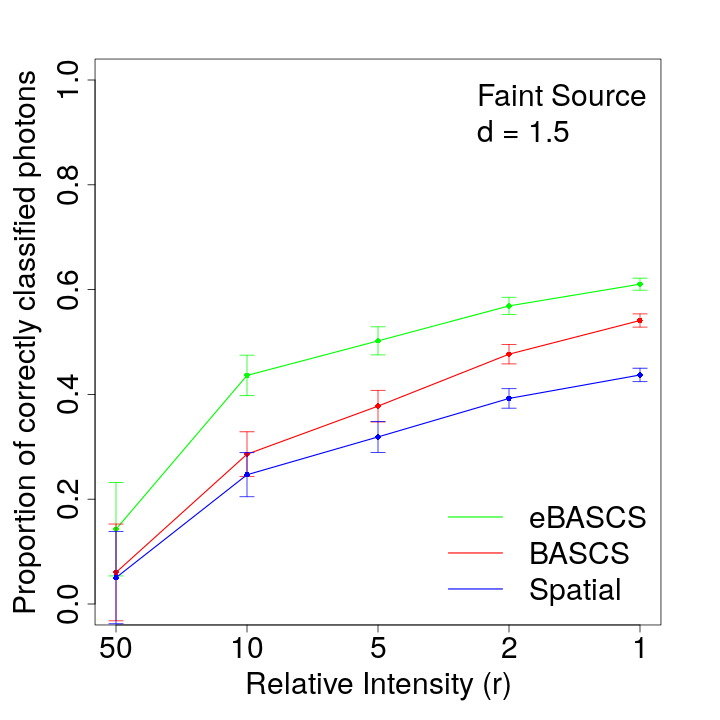}
    \caption{Average proportion of events originating from the faint source that were correctly classified by the \ebascs{} (green), \bascs{} (red) and \spatial{}  (blue) algorithms, for simulation settings $(d = 1.5, r \in \{1,2,5,10,50\})$, averaged over the replicate data sets. This plot is replicated for the other simulation studies in Appendix~\ref{sec:perfeval}. 
    }
    \label{fig:perfs2}
\end{figure}


\section{Simulation Studies} 
\label{sec:simulation_studies}

In this section, we evaluate the benefit of incorporating temporal data, 
by comparing the performance of \ebascs{} with the \bascs{} and \spatial{} algorithms on simulated data sets. Our simulations are parametrized in terms of a range of background intensities, spatial source separations, and relative source intensities. They are organized into two simulation studies that differ in terms of the strength of the background.
Simulation~I, described in Section~\ref{sec:highback}, constructs a challenging scenario where one source is weak compared to the background contamination. Simulation~II, described in Section~\ref{s:lowbgsim}, considers a more realistic situation where the background is weaker than the sources, in order to mimic the level of background encountered with modern high-resolution X-ray telescopes like \chandra. Section~\ref{s:simdesign} sets up the general simulation design, which applies to both Simulations~I and~II.


\subsection{Simulation Design}\label{s:simdesign}

Our simulation design involves two astronomical objects, one brighter than the other, emitting photons in an image of 10 by 10 spatial units\footnote{In terms of the 2D King profile PSF parametrized as in Appendix~\ref{appsec:king}, the 10 by 10 image
is roughly equivalent to the region where the PSF is greater than $5 \times 10^{-4}$.}. 
The first simulation setting is the distance between the two sources, denoted as $d$, and we consider the set of values $d \in \{0.5,1,1.5,2\}$. In each simulation, the number of events originating from the background and sources are drawn from Poisson distributions with respective means $m_0$, $m_1$ and $m_2$. The brighter source has $m_1 = 2000$, while the fainter source's intensity is defined as $m_2 = m_1  / r$, where $r$ denotes the relative intensity of the two sources and is the second simulation setting. We consider the set of  values  $r \in \{1,2,5,10,50\}$. The strength of the simulated background $m_0$ differs between Simulation~I and~II, see  Sections \ref{sec:highback} and \ref{s:lowbgsim} for details.


We also generate 
spectral data for the source and background events. Since the focus of Simulations~I and~II is to investigate how much we can improve the fitted parameters and event allocations by incorporating temporal data, we use a common spectra for both sources (rather than different spectra between the sources or between the time bins).
Specifically, 
the two sources 
have a common spectra generated from a Gamma distribution with shape parameter $\alpha = 3.18$ and mean $\gamma = 1832$; these particular parameter values follow \citet{2015ApJ...808..137J}. The background spectra is generated from a Uniform distribution.

To simulate the temporal data (in the form of arrival times for the events), a 60ks observation period is defined and split into four equally sized time bins. Events are allocated to each time bin according to a Multinomial distribution with parameters $(\lambda_1, \lambda_2,\lambda_3,\lambda_4) = (0.05,0.15,0.3,0.5)$ for the bright source, and $(\lambda_1, \lambda_2,\lambda_3,\lambda_4) = (0.5,0.3,0.15,0.05)$ for the faint source. This means that the bright source gets brighter over time, and that the faint source dims over the observation period (see right panel of Figure \ref{fig:simsetup}). Background events are spread uniformly among the time bins. The number of parameters to be fitted for \ebascs{} grows linearly with the number of time bins. Simulating data and fitting the model with four time bins is a reasonable option to both generate enough discriminatory temporal information and moderate the required computational complexity.

In Simulation~I, replicate data sets were generated under 20 different settings, crossing $r\in\{1,2,5,10,50\}$ and $d\in\{0.5,1,1.5,2\}$. In Simulation~II, 5 different settings were considered, crossing $r\in\{1,2,5,10,50\}$ and $d=1$.
Evaluating the performance of \ebascs{} in this way ensures its consistency with the expected behaviour of a disentangling model in the physical environments represented by the datasets. 
In particular, the two sources are expected to be increasingly distinguishable as their spatial separation, $d$, grows. Similarly, as $r$ gets large the model is expected to easily detect the brighter source but detection of the dimmer source may remain difficult. 
Fifty replicate data sets were generated for each of the 20 settings in Simulation~I, and each of \ebascs{}, \bascs{} and \spatial{}  were run on each replicate. For Simulation~II, fifty replicate data sets were generated for each of the 5 settings, and each of \ebascs{}, \spacetime{}, \bascs{} and \spatial{} were run on each replicate. 

\subsection{Simulation~I: High Background}\label{sec:highback}

In Simulation~I, the strength of the simulated background $m_0$ is defined as a function of the faint source region and intensity. 
Specifically, we set $m_0$ so that the expected number of background and faint source counts is equal in the faint source region (defined as the area where the PSF is greater than 3$\%$ of its maximum). Mathematically, we let $q$ be the probability that a event from the faint source falls within this source region and set
$m_0 = q m_2$. This results in very intense background that strongly overwhelms the fainter source, allowing us to investigate to what extent  \ebascs{} outperforms \bascs{} and \spatial{} in an extremely noisy environment.

At each MCMC iteration, updated allocations of the events to sources or background are drawn from a Multinomial distribution with probabilities given by the posterior distribution for the variable $s$, computed with the latest sampled parameter values. From these allocations, two metrics are computed to measure the classification performance of \ebascs{}. Specifically, we report 
\begin{description}
\itemsep = 5pt 
\item[\emph{Allocation Recovery}:] The proportion of events originating from a source that were indeed allocated to that source.
\item[\emph{Allocation Accuracy}:] The proportion of events allocated to a source that actually originate from this source.
\end{description}
All proportions are averaged over both MCMC iterations and replicate data sets.
Averaging over the replicate data sets
reduces sampling variability, which can be substantial in low-count simulation settings
(between 2000 and 4000 counts with $r\in\{50,10,5\}$). Relying on a single replicate data set  might not accurately reflect the relative performance of the methods. Tables~\ref{tab:agivens} and \ref{tab:sgivena} report the two metrics, respectively, averaging over the 20 simulation settings. Appendices~\ref{sec:appagivens} and \ref{sec:appsgivena} report complete results for each simulation setting separately.  

As expected, on average \ebascs{} performs substantially better at disentangling overlapping sources than either \bascs{} or \spatial{}. This provides strong evidence of the ability of the proposed method to leverage the temporal information to extract discriminatory features. Tables~\ref{tab:agivens} and \ref{tab:sgivena} show that all models are less able to properly allocating events to the faint source than to the bright source. This is a direct consequence of the simulation design. 
Under the simulation settings where $r$ is large ($r \geq 5$),
the faint source is 
completely overwhelmed by the strong background. For example, with a relative intensity of $r=5$, the expected number of faint source and background events are $m_2 = 400$ and $m_0 \approx 1920$, respectively, while the expected number of bright source events is $m_1 = 2000$.

\begin{table}
    \centering
    \begin{tabular}{|llll|}
    \hline
     True Source & \ebascs{} & \bascs{} & \spatial{}\\
     \hline
     Bright Source & 0.771 & 0.723 & 0.651 \\
    Faint Source & 0.420 & 0.307 & 0.252 \\
    Background & 0.831 & 0.822 & 0.722 \\
    \hline
    \end{tabular}
\caption{Allocation recovery for each source by \ebascs{}, \bascs{}, and \spatial{}. The proportions are averaged over MCMC iterates, replicate data sets and simulation settings. A breakdown according to simulation settings appears in Appendix~\ref{sec:appagivens}.}
\label{tab:agivens}
\end{table}
\begin{table}
    \centering
    \begin{tabular}{|llll|}
    \hline
     Allocated Source & \ebascs{} & \bascs{} & \spatial{}\\
     \hline
    Bright Source & 0.798 & 0.744 & 0.686 \\
    Faint Source & 0.424 & 0.310 & 0.264 \\
    Background & 0.803 & 0.794 & 0.671 \\
    \hline
    \end{tabular}
\caption{As in Table~\ref{tab:agivens}, for allocation accuracy. A breakdown according to simulation settings appears in Appendix~\ref{sec:appsgivena}}

\label{tab:sgivena}
\end{table}

\begin{figure*}
    \centering
    \includegraphics[width = 0.32\linewidth]{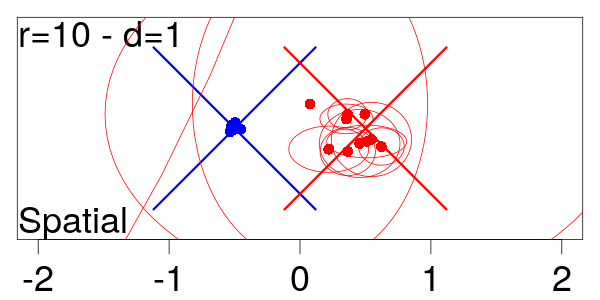}
    \includegraphics[width = 0.32\linewidth]{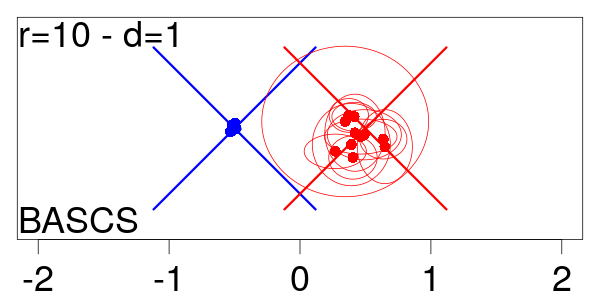}
    \includegraphics[width = 0.32\linewidth]{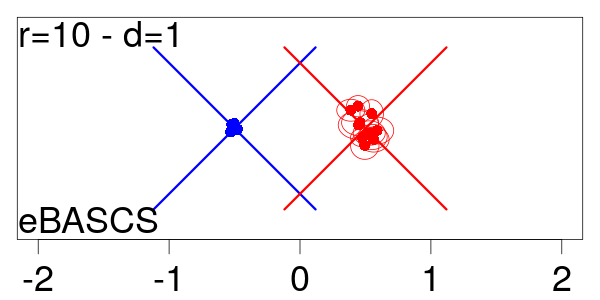}
    \caption{Posterior means and standard deviations of the source locations for 10 replicate data sets generated under the simulation setting with $r = 10$ and $d = 1$ and fit with the \spatial{}  (Left Panel), \bascs{} (Middle Panel) and \ebascs{} (Right Panel) algorithms. The points and ellipses represent the posterior means and standard deviations
    for each of the ten replicates; those for the bright source are plotted in blue and those for the faint source in red. 
    Crosses indicate the true location of the sources. \ebascs{} is able to locate the faint source much more consistently (the posterior means of the location are closer to the true value) and confidently (the posterior standard deviations are much smaller) than the other methods. 
    }
    \label{fig:senslocstd}
\end{figure*}


Figure~\ref{fig:perfs2} shows that classification performance for the faint source improves as its relative intensity grows (i.e., as $r$ decreases). 
In the simulation setting with $r = 1$ and $d = 0.5$, \ebascs{} had an allocation recovery of $57.1\%$ of the bright source events and $56.4\%$ of faint source events (see Table~\ref{tab:a1s1eB} in Appendix~\ref{sec:appagivens}). That is, in a setting where the sources are extremely close, have the same spectra and relative intensity, and are immersed in an exceedingly strong background, \ebascs{} still correctly classifies more than half of events from the sources. This is a substantial improvement over \bascs{}, which had an allocation recovery of only $42.8\%$ of bright source events and $42.2\%$ of faint source events in the same simulation setting (see Tables~\ref{tab:a1s1rb} and \ref{tab:a1s1sp} in Appendix~\ref{sec:appagivens}). 

The biggest improvement of \ebascs{} over the other algorithms is its ability to distinguish 
faint sources from background, particularly when the relative intensities of the bright and faint sources are extreme. In 
such a case, 
\bascs{} and \spatial{} sometimes mistake the faint source for a random cluster of events from either the brighter source or the background. In the middle panel of Figure \ref{tab:appperfeval0.5} in Appendix \ref{sec:perfeval}, for the simulation setting $r=10$ the allocation recovery of faint source events by \bascs{} and \spatial{} is close to 0, which indicates that these algorithms cannot separate the faint source from the bright source or the background. \ebascs{}, however, is able to better distinguish the faint source and has an allocation recovery of around 20$\%$.
Incorporating the temporal data allows \ebascs{} to partially solve this problem, and to provide more precise and confident event allocations.


In addition to probabilistically attributing events to sources, \ebascs{} provides estimates and error bars for the model parameters. Figures~\ref{tab:sensloc} and \ref{tab:sensloc2} (in Appendix~\ref{sec:appsensloc}) illustrate the statistical properties of the estimates by plotting the posterior means 
of the two source locations for all 50 replicate data sets under each of the 20 simulation setting.
Crosses indicate the true source locations.
The variance among replicate data sets of
the 
posterior mean of the
bright source's location increases as $r$ decreases, i.e., as the faint source becomes relatively brighter and the background intensifies compared to both sources. (Recall that the background intensity is tied to the faint source intensity.) The location of the faint source, however, is estimated more accurately as it grows brighter (i.e., as $r$ decreases). Particularly, when it is bright enough for the algorithms to detect an  energy signature that is distinguishable from the background spectrum.

The \ebascs{} algorithm yields fits that are 
at least as
accurate as \bascs{}. Even when the sources are closely located (i.e., $d=0.5$), the \ebascs{}-fitted posterior 
means of the source
locations concentrate 
more closely
around the true locations than those provided by \bascs{}.
With $d=0.5$, the separation between sources and background relies heavily on the energy model and
the \spatial{} algorithm does not perform as well as either \bascs{} or \ebascs{}.

The specific simulation setting with $r=10$ and $d=1$ is illustrated in Figure~\ref{fig:senslocstd} and shows that  \ebascs{} is not only able to locate the sources more accurately, as its posterior means  cluster more closer to the true location, but also more confidently, since the \ebascs{} standard deviations (circling the posterior means in Figure~\ref{fig:senslocstd}) are much smaller than those of \bascs{} and \spatial{}. 
This also holds for the other model parameters, see Appendix~\ref{sec:appsimulationpara} for a full comparison of the parameter estimates.

To investigate the number of counts \ebascs{} requires to produce meaningful results, we repeated the simulation setting illustrated in Figure~\ref{fig:senslocstd} ($r = 10$, $d=1$) but with $m_1 = \{800,600,400\}$; recall that $m_1$ is the expected count for the brighter source and was originally set to $m_1 = 2000$ in Simulation~I. Even with $m_1 = 400$, corresponding to about 100 counts per time bin, the posterior means of the source locations cluster closely around their true values.

\subsection{Simulation~II: Low Background}\label{s:lowbgsim}
The background intensities in Simulation~I are by design the same as the intensity of the faint source. 
%
The background in modern high-resolution X-ray telescopes similar to \chandra, however, is 
typically
weaker than this. Simulation~II assesses how \ebascs\ performs in a more realistic and plausible noise environment. It considers scenarios where the total background count is drawn from a Poisson distribution with mean 
$m_0=100$.

 \begin{figure}
    \centering
    \includegraphics[width = \linewidth]{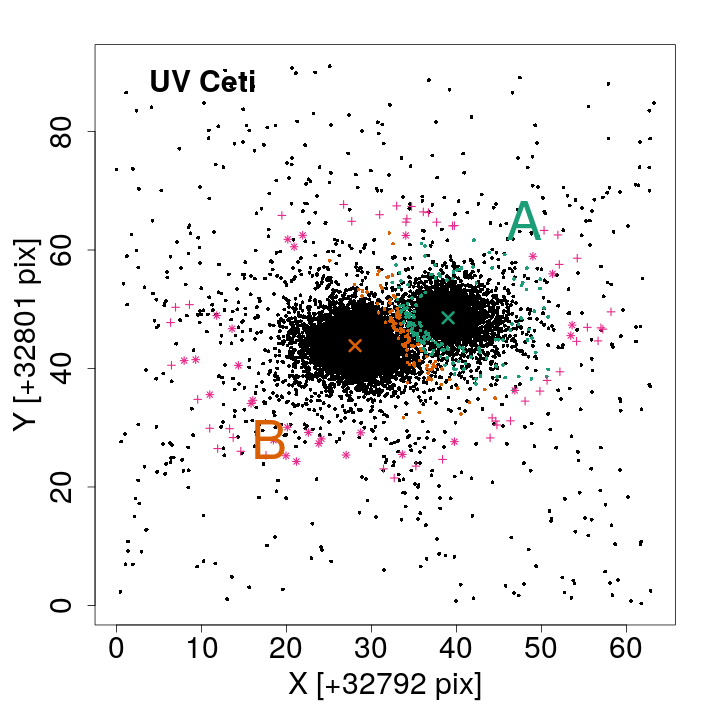}
    \caption{
    Sky pixel locations of photon
    events disputed by the \spatial{} and \spacetime{} algorithms for \uvcet. 
    Events coloured in black were allocated to the same source by both algorithms. Events coloured in orange were allocated to \uvcet{~B} by the \spatial{} algorithm and \uvcet{~A} by the \spacetime{} algorithm.  Events coloured in green are the
    opposite, i.e., allocated to \uvcet{~A} by \spatial\ but to \uvcet{~B} by \spacetime.  
    Disputed background events are marked as magenta symbols (``+'' denoting those reallocated to a source, and ``*'' are those reallocated to the background by \spacetime).  
    }
    \label{fig:diffspace}
\end{figure}

The allocation recovery and allocation accuracy for each algorithm and setting in Simulation~II are reported in Appendix~\ref{sec:applowback}. Tables~\ref{tab:lowsimres2} and \ref{tab:lowsimres1} average these metrics across relative source intensities, $r$, to summarize the overall performance of the algorithms in Simulation~II. 
The fact that the \spacetime{} algorithm outperforms \bascs{} for both sources is not unexpected. Because both sources are simulated with the same spectrum, the spectral data do not help distinguish between the two sources (but do help to separate out background events).  Temporal data, however, do help to distinguish the two sources. 
Since background counts are low in 
Simulation~II,
the added benefit of spectral data is small compared to that of temporal data, as illustrated by the almost imperceptible improvement of \bascs{} over the \spatial{} algorithm.

\begin{table}
\centering
\begin{tabular}{|lllll|}
\hline
  True Source &  \ebascs{} & \bascs{} & \spacetime{} & \spatial{}\\
\hline
  Bright Source & 0.869 & 0.804 & 0.841 & 0.810 \\
Faint Source & 0.495 & 0.346 & 0.499 & 0.334 \\
  Background & 0.580 & 0.570 & 0.315 & 0.200 \\
\hline
\end{tabular}
\caption{Allocation recovery. As in Table \ref{tab:agivens}, for Simulation~II by \ebascs{}, \bascs{}, \spacetime{} and \spatial{}. A breakdown according to simulation settings appears in Appendix~\ref{subsec:lowbackagivens}. 
}
\label{tab:lowsimres2}
\end{table}
\begin{table}
\centering
\begin{tabular}{|lllll|}
\hline
  Allocated Source &  \ebascs{} & \bascs{} & \spacetime{} & \spatial{}\\
\hline
  Bright Source & 0.878 & 0.820 & 0.870 & 0.780 \\
Faint Source & 0.484 & 0.334 & 0.338 & 0.482 \\
  Background & 0.523 & 0.515 & 0.305 & 0.221 \\
\hline
\end{tabular}
\caption{Allocation accuracy. As in Table~\ref{tab:lowsimres2}, for allocation accuracy.
A breakdown according to simulation settings appears in Appendix~\ref{subsec:lowbacksgivena}.}
\label{tab:lowsimres1}
\end{table}

\section{Application~I: UV Ceti}
\label{sec:uvcet}

\subsection{Data and Models}

UV\,Ceti (Gliese 65) is an M dwarf hierarchical binary system.  Both the main components of the binary, \uvcet\,A and \uvcet\,B are flare stars that undergo unpredictable and dramatic changes in their brightness over short timescales. 
The \uvcet\ system was observed with \chandra\ on 2001~Nov~26 (ObsID 1880) with the LETGS+HRC-S
configuration.  
The spatial resolution of \chandra\ is sufficient to visually distinguish the two components, which are separated by a distance of 1.4~arcsec (\uvcet\,B is itself a binary \citep{1998A&A...331..596B}, but with a separation of $\sim$1~mas, it is not resolvable by \chandra).
These data have been previously analyzed to separate the two components \citep[see][]{2003ApJ...589..983A} using small extraction radii to limit contamination of one source due to the other.  However, \uvcet\,B undergoes a large flare during the observation, and its effect is also visible in the light curve of \uvcet\,A \citep[see Fig.~2]{2003ApJ...589..983A}.  This source is thus a natural test case for the application of the \ebascs\ algorithm.  Since the HRC-S has no spectral discrimination, applying \bascs\ to the $0^{\rm th}$-order data to separate the events from the two sources will not yield any improvements over the \spatial\ algorithm.  We demonstrate below that the \spacetime\ algorithm does lead to a significantly better allocation of events.

\begin{figure*}
    \centering
    \includegraphics[width = 0.45\linewidth]{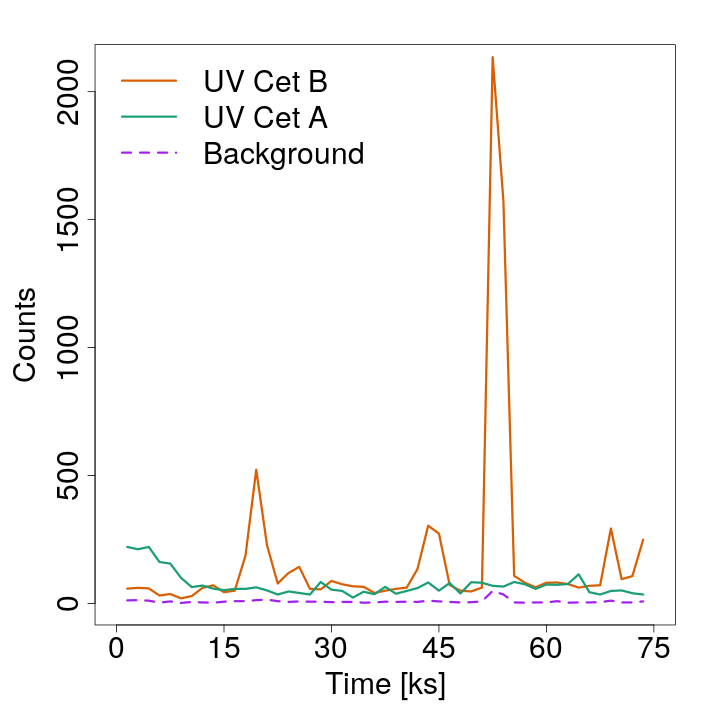}
    \includegraphics[width = 0.45\linewidth]{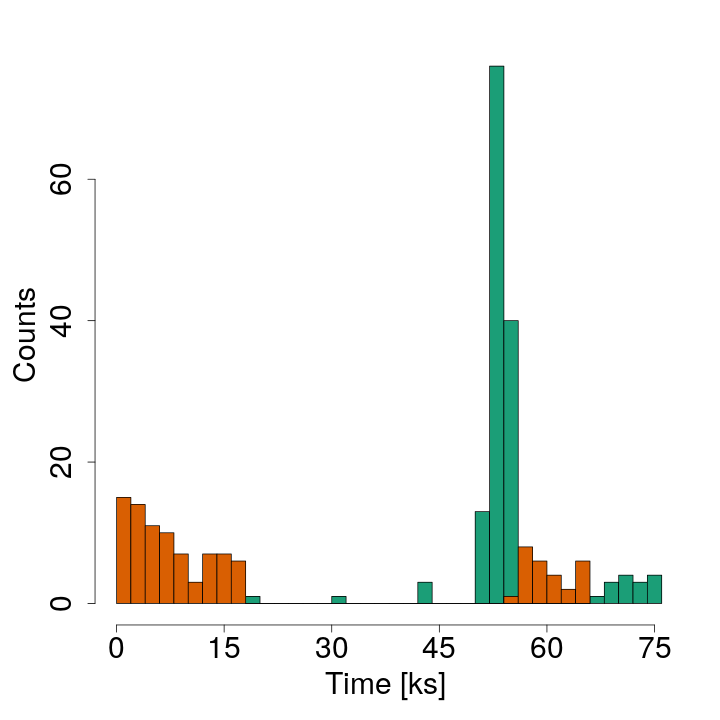}
    \caption{\emph{Left Panel:} Light curves for UV Cet A, UV Cet B and the background obtained 
    from an single Monte Carlo iteration of the
    \spacetime{} 
    algorithm. There is moderate contamination from the 
    \uvcet\,B's flare 
    (at around 53ks) 
    on the background. This 
    contamination is less severe than with the 
    \spatial\  algorithm; \spacetime{} allocated 84 events to the background at the time of the \uvcet\,B flare, while \spatial{} allocated 106 events to the background. (Such contamination is also a consequence of the approximate nature of our PSF model.)
    \emph{Right Panel:} Arrival times
    of events disputed by the \spatial{} and \spacetime{} algorithms. The orange bars indicate events that are moved from \uvcet\,B to \uvcet\,A, and green bars indicate events that are moved from \uvcet\,A to \uvcet\,B by the \spacetime{} model.
    Events allocated to the same source by both algorithms are not included in the plot.  
    Notice that \spacetime\ is successful in identifying the contamination in \uvcet\,A due to the large flare of \uvcet\,B and allocating those events to \uvcet\,B.
    }
    \label{fig:difftime}
\end{figure*}

We selected the time bins for \uvcet\ using the procedure detailed in Section \ref{sub:binselection}. We chose 6 time bins which renders a temporal model that is flexible enough to capture the source flare at $\approx$54~ks after the start of observation
(see Figure \ref{fig:algoseg}) while maintaining a reasonable number of model parameters to be fit with the \spacetime{} algorithm (25 parameters i.e., 4 location, 3 relative intensity and 18 time intensity parameters).

\subsection{Results}
\label{sec:0th}

To measure the difference of the \spacetime{} model over the \spatial{} model, we carry out a 
disputed event analysis.
This consists of 
comparing the allocations of the events in the two algorithms
and studying the characteristics of disputed events
to highlight the benefits of incorporating temporal information.
For each event $i$, the disentangling models output the posterior distribution of $s_i$, the latent variable that encodes its origin. For each model, we set the allocation of event $i$ to the source it has the highest probability of originating from, i.e., event $i$ is allocated to the bright source if $p(s_i = \text{bright})  = \max\{p(s_i = \text{bright}), p(s_i = \text{faint}),p(s_i = \text{background})\}$. 

Figure \ref{fig:diffspace} shows that, as expected, most disputed events are located in the zone where the source wings overlap, or in other words where events are roughly equidistant from both cores. In this case, the \spacetime{} algorithm performs a more careful analysis of events in the overlap between the sources, i.e., in locations where it is most difficult to separate the sources based on spatial data alone.

\begin{figure*}
    \centering
    \includegraphics[width = 0.32\linewidth]{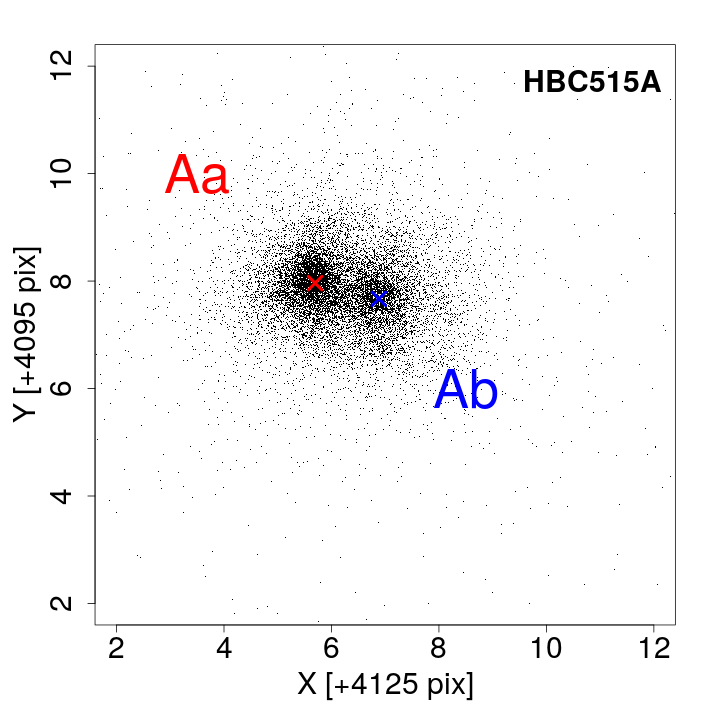}
    \includegraphics[width = 0.32\linewidth]{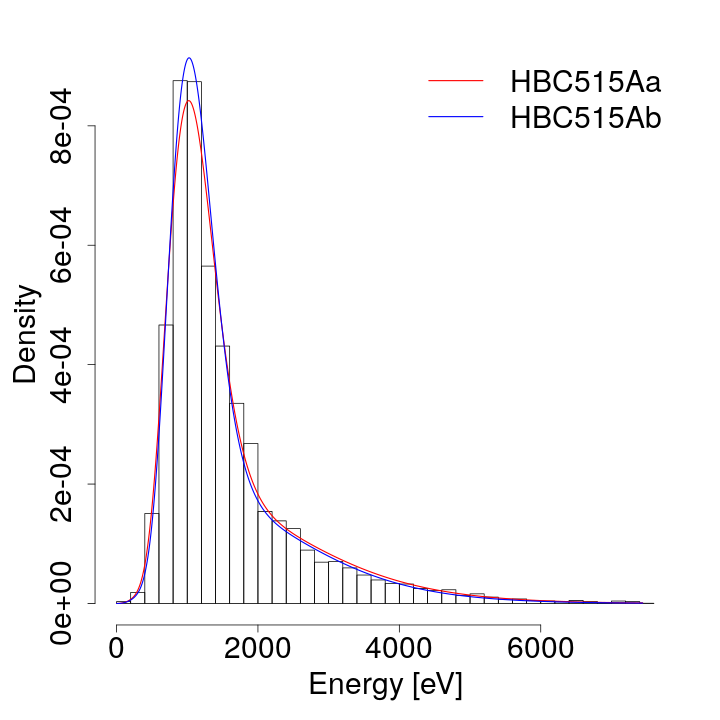}
    \includegraphics[width = 0.32\linewidth]{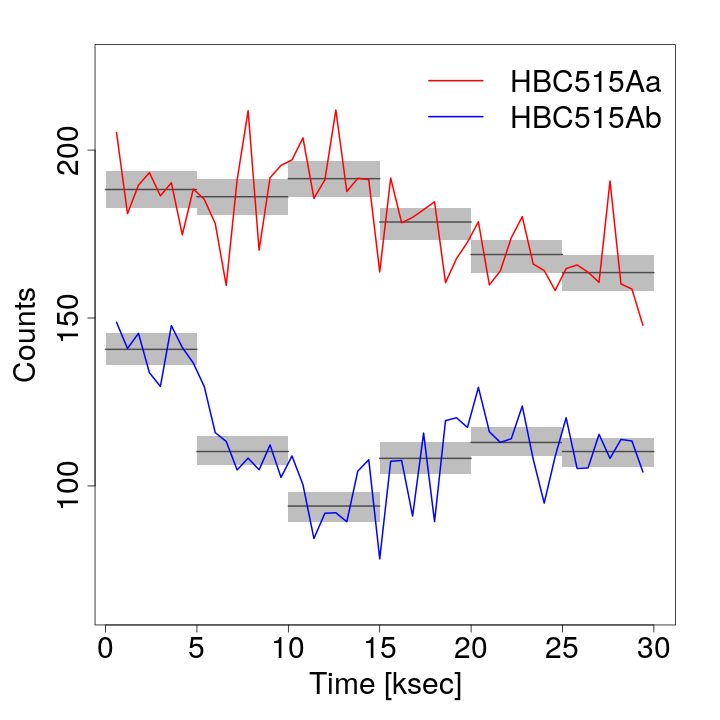}
    \caption{$\chandra$ ACIS-S observation of the binary system \hbc\,A. \emph{Left Panel:} Scatter plot 
    of recorded event locations with the posterior means of the locations of 
    \hbc~Aa and \hbc~Ab computed with \bascs{} plotted as red and blue crosses, respectively.
    \emph{Middle Panel:} Histogram of spectral data with 
    source spectra superimposed (using the posterior mean of the spectral parameters fitted with \bascs{})
    . \emph{Right Panel:} Light curves for the two separated components of \hbc\,A obtained by allocations made by \ebascs{}, superimposed with the \ebascs{}-fitted light curves (grey). The red and blue curves denote the average of the allocated light curves over 500 iterations of \ebascs{}, the dark gray horizontal lines denote the posterior mean of the temporal parameters fitted by \ebascs{}, and the light grey regions denote the intervals between the 16$\%$ and 84$\%$ posterior quantiles (see Table \ref{tab:params2}).}
    \label{fig:hbc515}
\end{figure*}

\begin{figure}
    \centering
    \includegraphics[width = \linewidth]{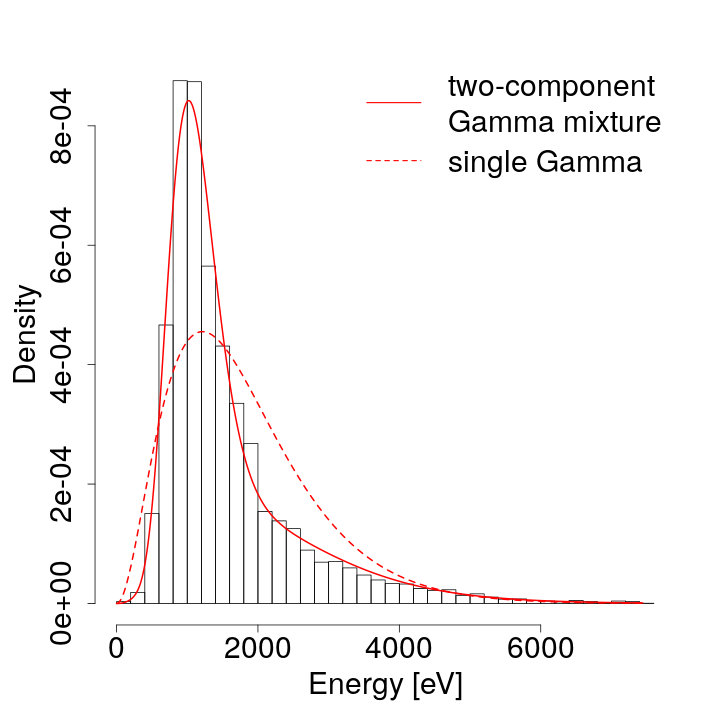}
    \caption{Fitting a single-Gamma distribution and a two-component mixture of Gammas distributions to the spectral data for \hbc\,Aa. The histogram shows the observed spectrum of the \hbc\,A system (i.e, all sources combined). Note that the middle panel of Figure \ref{fig:hbc515} indicates that \hbc\,Aa and \hbc\,Ab have very similar spectra. The solid line is the fitted two-component mixture of Gammas and the dashed line is fitted the single Gamma model.}
    \label{fig:specmodel}
\end{figure}

The 
recorded locations
of the disputed events allocated to \uvcet\,A by \spatial{}, circling around the source's core, indicate that these events were assigned to \uvcet\,B by \spacetime{} on the basis on temporal information. Indeed, as illustrated in the right panel Figure~\ref{fig:difftime}, these events were all detected at the time coincident with an observed flare attributed to \uvcet\,B. The right panel shows that this contamination of the flare of \uvcet\,B on \uvcet\,A is completely removed by \ebascs{}, as the lightcurve of \uvcet\,A (as allocated by \ebascs{}) does not exhibit a spike in its intensity at the time of the flare. Figure~\ref{fig:difftime} also shows that the contamination of \uvcet\,A on \uvcet\,B at early times (between 0ks and approximately 15ks), i.e., when \uvcet\,A has higher intensity, is removed by \ebascs{}.

\begin{table}
    \centering
    \begin{tabular}{|cccc|}
    \hline
    & \uvcet\,B & \uvcet\,A & Background \\
    \hline
    \uvcet\,B & 8388 & 149 & 28 \\
    \uvcet\,A & 107 & 3560 & 19 \\
    Background & 23 & 5 & 381 \\
    \hline
    \end{tabular}
    \caption{Disagreement matrix between the allocations made by the  \spacetime{} and \spatial{} algorithms. Columns correspond to
    the allocations made by \spatial{} and rows correspond to allocations  made  by \spacetime{}. For example, 149 events that were attributed to \uvcet\,A by \spatial{} were instead allocated to \uvcet\,B by \spacetime{}.}
    \label{tab:disagreement}
\end{table}

Table~\ref{tab:disagreement} reports the disagreement matrix (i.e., the number of events allocated to each source by both algorithms). 
Adding temporal data changed the allocation
of 331 (out of 12,660) events, 149 of which correspond to the contamination of \uvcet\,B's flare on \uvcet\,A, and 107 of which correspond to the contamination of \uvcet\,A on \uvcet\,B in the early stages of the observation period. Removing these contaminations would not have been possible without modelling the temporal information, and clearly shows the improvement of \spacetime{} over the \spatial{} algorithm.

\section{Application II: \hbc\,A}
\label{sec:hbc515}

\subsection{Data and Models}
\label{sec:hbc515d&m}
\hbc\,A is a component of {the} well-separated weak-lined T Tauri multi-component system HBC\,515 \citep{2010AJ....139.1668R}. \hbc\,A is a binary, composed of two variable stars, \hbc\,Aa and \hbc\,Ab,  separated by a distance of approximately 0.5 arcsec. 
The system was observed with \chandra/ACIS-S on 2011~January~8 (ObsID 12383).  
The overlap of the PSFs of the two components is significantly larger than \uvcet,
and separating them through non-overlapping extraction regions would lead to a significant loss in the number of counts available for follow-up analyses.  Such an analysis was performed by \citet{2017A&A...598A...8P}, who used a complex set of regions covering the cores and the diametrically-opposed crescent-shaped wings to extract the events and conduct spectral and temporal analyses.  They found that the spectra accumulated over the duration of the observation were similar, and that there was no evidence for temporal variability.
In the following, we demonstrate that even in this challenging dataset, \ebascs\ is able to recover small temporal changes to both the intensities and spectra of the two components. 
Not only do the data show that the spectra of both stars vary stochastically over timescales of a few ks, but that at different times different components are observed to be {spectrally} harder.

\begin{figure*}
    \centering
    \includegraphics[width = 0.88\linewidth]{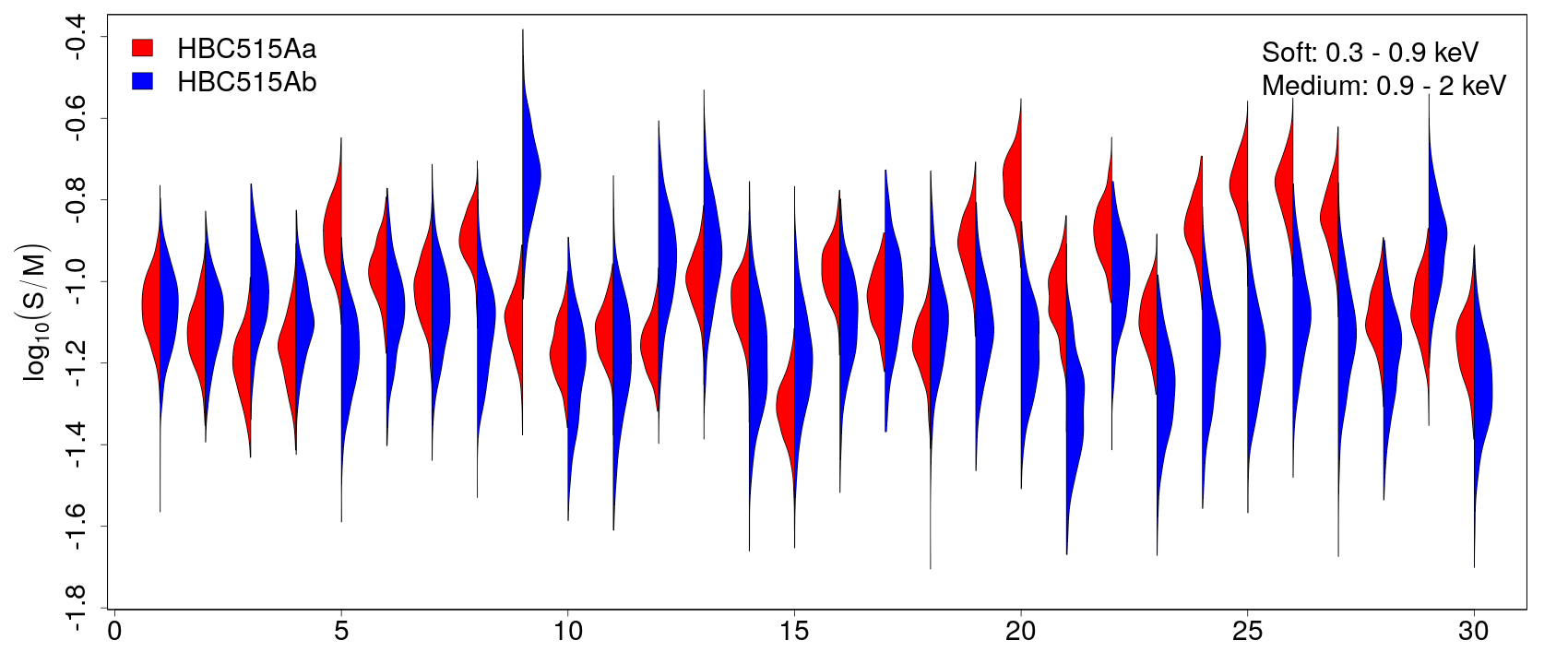}
    \includegraphics[width = 0.88\linewidth]{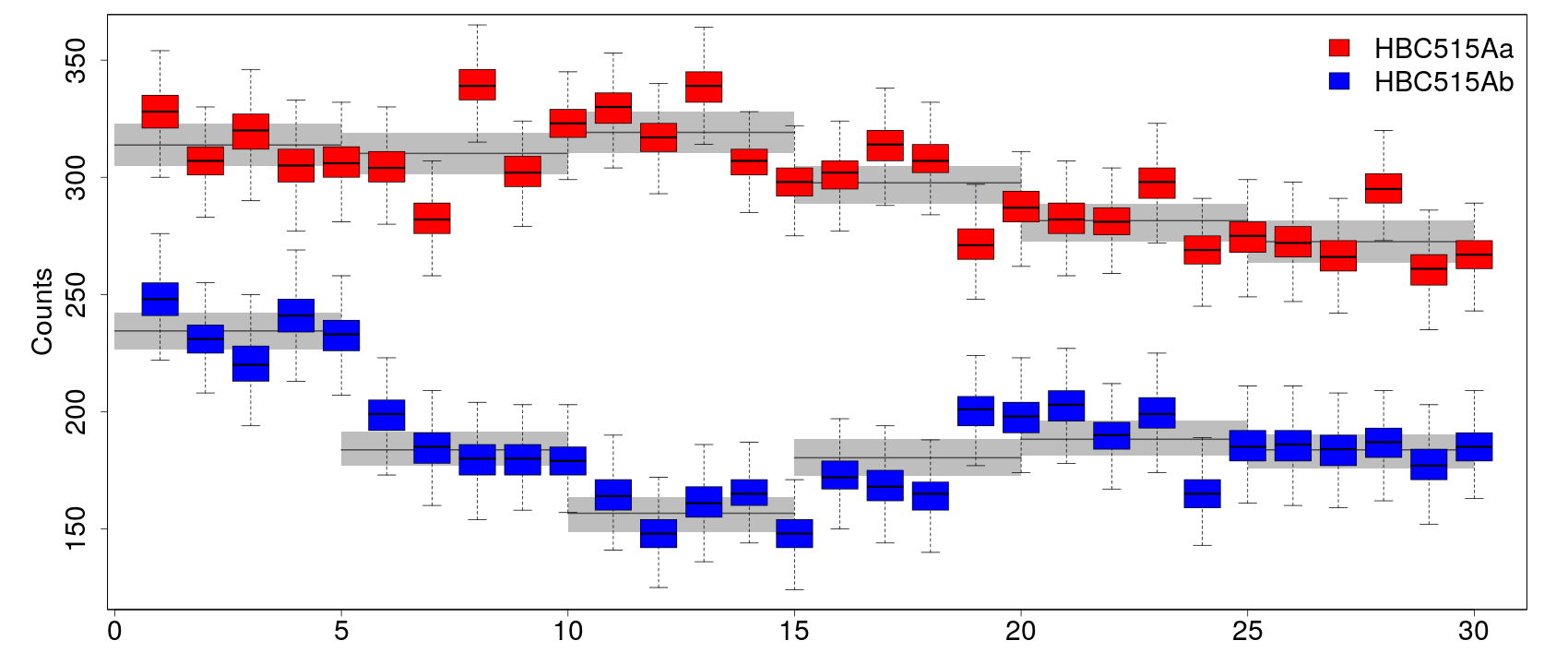}
    \includegraphics[width = 0.88\linewidth]{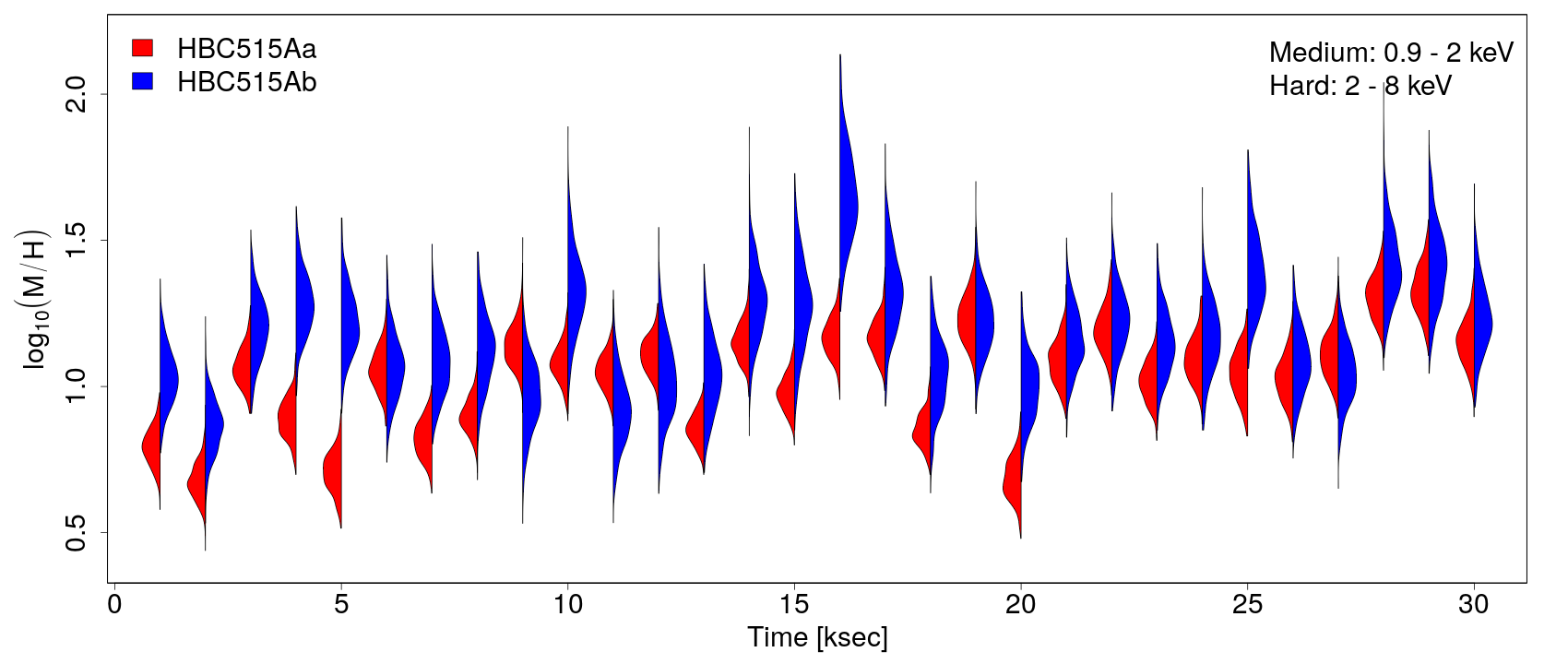}
    \caption{
    Spectral and intensity variability in the components of \hbc\,A, constructed using event allocations from 500 iterations of \ebascs.  {\sl Top and Bottom Panels:}
    Spectral hardness in the Soft ($S$:0.3-0.9~keV), Medium ($M$:0.9-2~keV), and Hard ($H$:2-8~keV) bands ({\sl Top:} C$_{SM}=\log\frac{S}{M}$; {\sl Bottom:} C$_{MH}=\log\frac{M}{H}$) are shown for events grouped into 1~ks bins for both the \hbc\,Aa (red) and \hbc\,Bb (blue) components.  The posterior distributions of spectral color are shown oriented vertically for each time bin.  These so-called violin plots are normalized so that the modes have the same horizontal displacements; thicker part of each color segment denote a higher posterior probability of the ordinate in a given time bin.  In both panels, the left halves of the shapes represent \hbc\,Aa, and the right halves represent \hbc\,Ab in order to facilitate comparisons. Counts for the individual violin plots for HBC 515Aa vary from 49-76 (Soft), 146-196 (Medium), and 43-85 (Hard); counts for HBC 515Ab vary from 28-50 (Soft), 84-150 (Medium), and 22-55 (Hard). {\sl Middle Panel:}
    Intensity variations in the combined broad ($B=S+M+H$:0.3-8~keV) passband are shown for the same time binning and demonstrate the evolution of the brightness of each component.  The fitted intensities $\lambda_{i,j}$ in each time segment (see Table~\ref{tab:SpaceTimeUVCet}) are shown as grey bands.
    Overall the panels illustrate several instances of spectral changes, e.g., \hbc\,Aa is spectrally harder at the beginning of the observation (see C$_{MH}$), and is spectrally softer near the 25~ks mark (see C$_{SM}$); \hbc\,Ab shows increasing spectral hardening as it recovers from 
    its minimum brightness.
    }
    \label{fig:hardness}
\end{figure*}

The \hbc\,A dataset is comprised of 14601 events recorded over 28.76~ks.  We first extract the events over a range of sky pixels large enough to include both components of the binary, but exclude other components of the system (see Figure~\ref{fig:hbc515}).\footnote{As a practical matter, in order to allow for better and faster convergence, it is important to ensure that the every part of the spatial window is covered by the wings of the PSF wherever it may be centered, as otherwise it is possible for one of the fitted source locations to stick in a random cluster of background events too distant from the sources' cores, or in a more event-sparse region of the image.}

Since the ACIS detectors do allow for event energy discrimination, we apply the \ebascs\ 
algorithm to the \hbc\,A data, allowing us to characterize the time variability of both \hbc\,Aa and \hbc\,Ab.  We use \ebascs\ with the number of sources fixed at $S=2$, and for simplicity adopt the King profile density (Appendix~\ref{appsec:king}) as the PSF model.  We use six equally-spaced time bins to model temporal variations, as this choice presents a practical trade off between preserving the flexibility in modeling while limiting the size of the parameter space.  We model the counts spectra in three different ways: first, using the single-Gamma distribution (Equation~\ref{eq:singlegamma}) 
for each source
; second, using the
two-component mixture of Gamma distributions (Equation~\ref{eq:mixgamma}) 
for each source; and third, using a two component mixture at each time bin 
for each source (Equation~\ref{eq:eBASCS_energy}).  As discussed by \citet{2015ApJ...808..137J}, a crude spectral model that roughly tracks the shape of the astrophysical model spectrum weighted by the effective area has sufficient power to distinguish spectral variations between the sources. 
We use a uniform spectral model for the background.

\subsection{Results}

The two-component mixture of Gamma distributions (Equation \ref{eq:mixgamma}) fits the observed counts spectra better than the single-Gamma distribution (see Figure \ref{fig:specmodel}).
Unfortunately, the very few counts in each time bin per source leads to large uncertainties in the parameter estimates  under the spectral model that incorporates a two component mixture at each time bin. Moreover, the spectra of the sources do not appear to be variable enough to justify the complexity of estimating different spectral parameters for each source at each time bin. 
Thus, we adopt the two-component mixtures for each source, but do not allow them to vary between time bins.

Table \ref{tab:params1} in Appendix \ref{sec:paramshbc515} shows that parameter values  fitted by \bascs{} are recovered by  \ebascs{} with similar precision. This demonstrates the consistency and well-definedness of  \ebascs{}. Both models return a posterior mean of the relative intensity of the background of $\approx$0; this is a consequence of 
the cropping of the image described in Section~ \ref{sec:hbc515d&m}.

Our estimates of the two-component Gamma mixture model parameters indicate that \hbc\,Aa and  \hbc\,Ab have very similar spectra (see middle panel of Figure \ref{fig:hbc515}), as \citet{2017A&A...598A...8P} suggested.

\begin{table}
    \small
    \centering
    \begin{tabular}{|rlll|}
    \hline
    & mean & q16 & q84\\
    \hline
    \multicolumn{1}{c}{\hbc\,Aa}\\
    $\lambda_{1,1}$ & 0.175 & 0.170 & 0.180\\
    $\lambda_{1,2}$ & 0.173 & 0.168 & 0.178\\
    $\lambda_{1,3}$ & 0.178 & 0.173 & 0.183\\
    $\lambda_{1,4}$ & 0.166 & 0.161 & 0.170\\
    $\lambda_{1,5}$ & 0.157 & 0.152 & 0.161\\
    $\lambda_{1,6}$ & 0.152 & 0.147 & 0.157\\
    \hline
    \multicolumn{1}{c}{\hbc\,Ab}\\
    $\lambda_{2,1}$ & 0.208 & 0.201 & 0.215\\
    $\lambda_{2,2}$ & 0.163 & 0.157 & 0.170\\
    $\lambda_{2,3}$ & 0.139 & 0.132 & 0.145\\
    $\lambda_{2,4}$ & 0.160 & 0.153 & 0.167\\
    $\lambda_{2,5}$ & 0.167 & 0.161 & 0.174\\
    $\lambda_{2,6}$ & 0.163 & 0.156 & 0.169\\
    \hline
    \multicolumn{1}{c}{Background}\\
    $\lambda_{3,1}$ & 0.170 & 0.036 & 0.313\\
    $\lambda_{3,2}$ & 0.168 & 0.034 & 0.310\\
    $\lambda_{3,3}$ & 0.164 & 0.034 & 0.300\\
    $\lambda_{3,4}$ & 0.168 & 0.034 & 0.310\\
    $\lambda_{3,5}$ & 0.167 & 0.035 & 0.306\\
    $\lambda_{3,6}$ & 0.164 & 0.034 & 0.302\\
    \hline
    \end{tabular}
    \caption{Temporal parameters of \hbc\,A fitted with \ebascs{}. $\lambda_{j,k}$ denotes the relative intensity of source $j$ in time bin $k$. Here $j=1$ corresponds to \hbc\,Aa, $j=2$ to \hbc\,Ab and $j=3$ to the background. The first column gives the posterior mean of the corresponding parameters, and columns "q16" and "q84" respectively denote their 16$\%$ and 84 $\%$ posterior quantiles.}
    \label{tab:params2}
\end{table}

\ebascs{} is able to recover temporal changes in the intensities of the two components (see Table \ref{tab:params2} and right panel of Figure \ref{fig:hbc515}). The estimated temporal parameters shown in Table \ref{tab:params2} reveal statistical evidence for a difference in the light curves of \hbc\,Aa and  \hbc\,Ab. Indeed, the error bars (computed using 16$\%$ and 84$\%$ posterior quantiles) of the \ebascs{}-fitted temporal parameters $\lambda_{1,k}$ and $\lambda_{2,k}$
do not significantly overlap in five of the six time bins. (Only error bars for $\lambda_{1,4}$ and $\lambda_{2,4}$ show a substantial overlap.) The right panel of Figure \ref{fig:hbc515} illustrates the difference in the separated lightcurves; \hbc\,Aa is stable for the first 15ks and then starts dimming, whereas \hbc\,Ab has a u-shaped light curve.

To further investigate  spectral differences between the sources, we analyse variations in their hardness ratios, shown in Figure~\ref{fig:hardness}. First, we sampled 500 allocations of the recorded events to \hbc\,Aa and \hbc\,Ab from the posterior distribution of $s$ (i.e., the latent variable encoding the origin of the observed events, see Section~\ref{sec:FiniteMix}) under \ebascs{}. Then, for each allocation, we computed the spectral hardness of the separated sources in the Soft ($S$:0.3-0.9~keV), Medium ($M$:0.9-2~keV), and Hard ($H$:2-8~keV) bands, in each of 30 time intervals of length 1ks. This yields, for each separated source at each time interval, the posterior distribution of spectral hardness in the top ($\log \frac{S}{M}$) and bottom ($\log \frac{M}{H}$) panels of Figure~\ref{fig:hardness}.
Figure~\ref{fig:hardness} shows that both sources exhibit variations in their spectra over the observation period. 
\ebascs{} is able to identify time scales over which \hbc\,Aa and \hbc\,Ab exhibit differences in their hardness ratios, see caption of Figure~\ref{fig:hardness} for details.

\section{Summary}\label{sec:conc}

We have presented \ebascs, an extension to the \bascs{} method developed by \citet{2015ApJ...808..137J} to leverage temporal variability signatures in high-energy astronomical sources with overlapping point spread functions to perform a better separation of the photon events.  The method integrates the temporal information into the disentangling algorithm via a flexible model which allows us to extract discriminatory features from the observed data. The assumption of independence of the brightness across time bins allows the model to flexibly capture temporal variability.

Several enhancements to \ebascs\ are in progress.  We plan to enhance the scalability of the method, while maintaining its current flexibility, by modelling the temporal information with simple continuous-time processes; incorporate instrument sensitivity and model the spectra using physically meaningful models for the source spectra; explore extensions of our spectral modelling to grating data (e.g., to separate photons in overlapping lines in the \chandra\ LETGS+HRC-S \uvcet\ observation); apply our methodology to astronomical systems that exhibit higher contrast in the relative intensities of their components (e.g., weak jets of X-ray bright quasars); explore observations from instruments with lower spatial resolution (such as \textit{NuSTAR}) to investigate whether \ebascs\ is able to separate spatially unresolved sources on the basis of their spectral and temporal variations; and finally, extend the method to allow the number of sources in the model to be estimated by carrying out both model comparisons for different assumed numbers of sources (e.g., using AIC \citep{1100705} or BIC \citep{10.1214/aos/1176344136}) as well as using a more sophisticated Reversible Jump MCMC method \citep{10.1093/biomet/82.4.711, 2015ApJ...808..137J}.



Simulation studies 
show 
that \ebascs{} achieves 
more accurate separation of photons from overlapping sources
than either \bascs{} 
or the baseline \spatial{} method. In particular, the proposed method further removes the contamination at the sources' cores and produces a 
better disambiguation of the
event allocation. \ebascs\ retains the advantages of performing inference under the Bayesian paradigm (i.e., uncertainty quantification, joint parameter inference, and probabilistic assignment of events) from its predecessor. 

The probabilistic allocation of events to sources can be incorporated in detailed follow-up spectral and temporal analyses. In particular, this uncertainty can be accounted for by repeatedly sampling event allocations from the posterior distribution of $s$, conducting the follow-up analysis according to the sampled allocations, and finally combining the results from each individual analysis.

Our application of \ebascs{} to the datasets \uvcet\ and \hbc\,A shows that our proposed model performs a more careful separation of the observed sources than other methods. The \spacetime{} model  almost eliminates the contamination of the flare of \uvcet\,B on \uvcet\,A. Applying \ebascs{} to the \hbc\,A data allows us to recover temporal changes in the intensities of \hbc\,Aa and \hbc\,Ab 
and to identify time intervals where the hardness ratios of the two components appear to differ.

Based on the Simulation results reported in Section~\ref{sec:simulation_studies} and the general statistical principle that including more data/information in a model yields more reliable estimates, we expect the overall performance of \ebascs{} to exceed that of the competing algorithms. Improvement is expected to be even more significant on systems with more substantial spatial overlap and distinct light curves among the sources.





\section*{Acknowledgements}
This work was conducted under the auspices of the CHASC International Astrostatistics Center. CHASC is supported by NSF DMS 15-13492, DMS 15-13484, DMS 15-13546, DMS-18-11308, DMS-18-11083, and DMS-18-11661, and by NASA grant
80-NSSC21-K0285.
We thank CHASC members for many helpful discussions, {especially Kathryn McKeough, Xiao-Li Meng, Alex Geringer-Sameth, and Chang Chuan Goh}. DvD and AZ were supported in part by a Marie-Skodowska-Curie RISE (H2020-MSCA-RISE-2015-691164, H2020-MSCA-RISE-2019-873089) Grants provided by the European Commission. VLK and AS acknowledge support from NASA Contract NAS8-03060 to the Chandra X-ray Center.

\section*{Data Availability}
The data used in this article are publicly available at the \chandra{} Data Archive (CDA) \footnote{\url{https://cxc.harvard.edu/cda/}}. The \bascs\ software is available on the CHASC GitHub software library\footnote{\url{https://github.com/AstroStat/}}, and \ebascs\ will also be made available at the same location.



\bibliographystyle{mnras}
\bibliography{eBASCS_final} 




\clearpage
\appendix

\section{King Profile Density}
\label{appsec:king}
The Point Spread Function used to generate the data in the simulation studies and fit the models (both in the simulations and dataset applications) is the 2D King profile, as in \citet{2015ApJ...808..137J}, whose functional form is given by:
\begin{equation}
    f(d) = \frac{C}{(1+(d/d_0)^2)^{\eta}}
\end{equation}
where
\begin{equation}
    d(x,y,w) = \sqrt{(x \cos w + y \sin w)^2 + (y \cos w - x \sin w)^2/(1-\epsilon)^2}
\end{equation}
The constant $C$ is determined numerically. The parameters are chosen to be: off-axis angle $\theta = 0$ arcmin, core radius $d_0 = 0.6$ arcsec, power-law slope $\eta = 1.5$, and ellipticity $\epsilon = 0.00574$.

\clearpage
\section{Time-Bin selection algorithm}
\label{sec:apptimebin}


The following time bin selection algorithms isolates transient variations and flares in the observed system light curve.  The algorithm initially 
evenly 
bins the temporal data,
into $\approx$50 bins,
then selects breakpoints in between which the largest variation in brightness occur, and finally applies a small deviation to the final breakpoints to avoid splitting the data too close the flares, for instance.
The parameters (to be chosen by the user) for this algorithm are:
\begin{itemize}
    \item the number of thin bins $k$ in the initial binning of the temporal data,
    \item the number of breakpoints $N$ to select,
    \item the deviation $d$ applied to the selected breakpoints.
\end{itemize}
The algorithm is designed as follows:

\begin{itemize}
    \item Thinly bin the observation period $[0,T]$ into $k$ bins.
    \item Count the number of observations $X_i$ in each time bin, and denote by $X = (X_1, X_2, \dots, X_k)$ the vector containing the counts in each bin.
    \item Compute the vector of first-order differences $ \Delta  X = (X_2 - X_1, \dots, X_k - X_{k-1})$.
    \item Identify the $N$ maximum values of $ \Delta  X$ and set them as breakpoints.
    \item If two breakpoints occur at consecutive bins, only keep one corresponding the higher value of $ \Delta  X$. If three breakpoints occur at consecutive bins, delete the middle one. This might cause the final number of breakpoints to be less than $N$, hence running the algorithm with different values of $N$ is recommended to produce a satisfactory selection.
    \item Denote by $\{b_1, \dots, b_N\}$ the selected breakpoints. Note that $\{b_1, \dots, b_N\}$ is a subset of $\Delta X$. For each $b_i$:
        \begin{itemize}
            \item If $b_i$ = $\Delta X_j$ for some $j$ such that $\Delta X_j \geq 0$, then set $b^{*}_i = b_i - d$.
            \item If $b_i$ = $\Delta X_j$ for some $j$ such that $\Delta X_j \leq 0$, then set $b^{*}_i = b_i + d$.
        \end{itemize}
    \item The algorithm returns $\{b^{*}_1, \dots, b^{*}_N\}$, the final breakpoints.
    
\end{itemize}

\onecolumn


\clearpage

\section{Simulation~I: Graphical comparison of proportion of correctly allocated events according to simulation settings.} \label{sec:perfeval}

\begin{figure}
   \begin{minipage}{0.3\linewidth}
     \centering
     \includegraphics[width=1\linewidth]{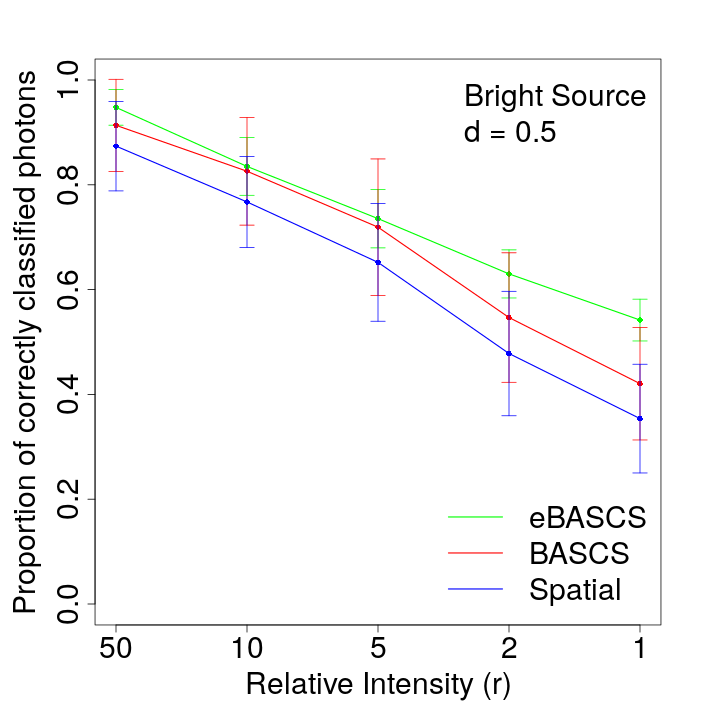}
   \end{minipage}\hfill
   \begin{minipage}{0.3\linewidth}
     \centering
     \includegraphics[width=1\linewidth]{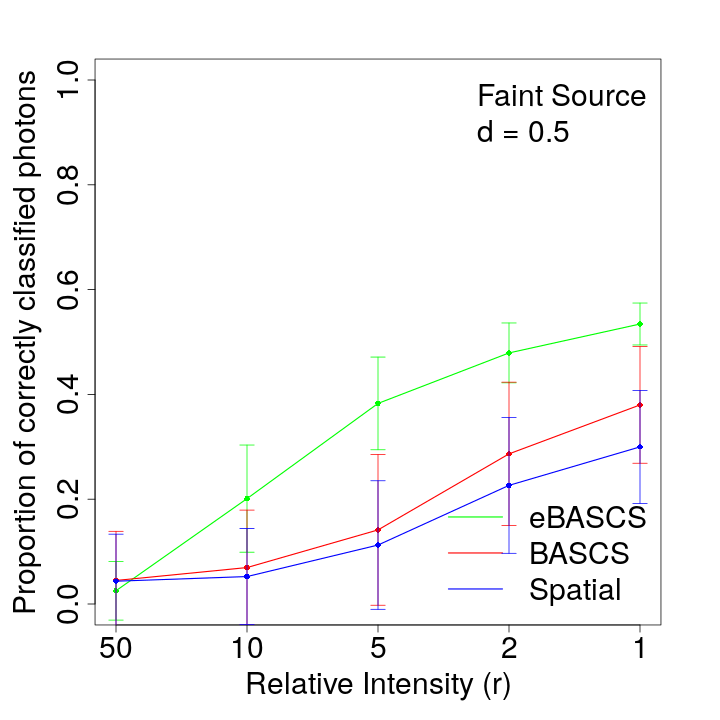}
   \end{minipage}\hfill
      \begin{minipage}{0.3\linewidth}
     \centering
     \includegraphics[width=1\linewidth]{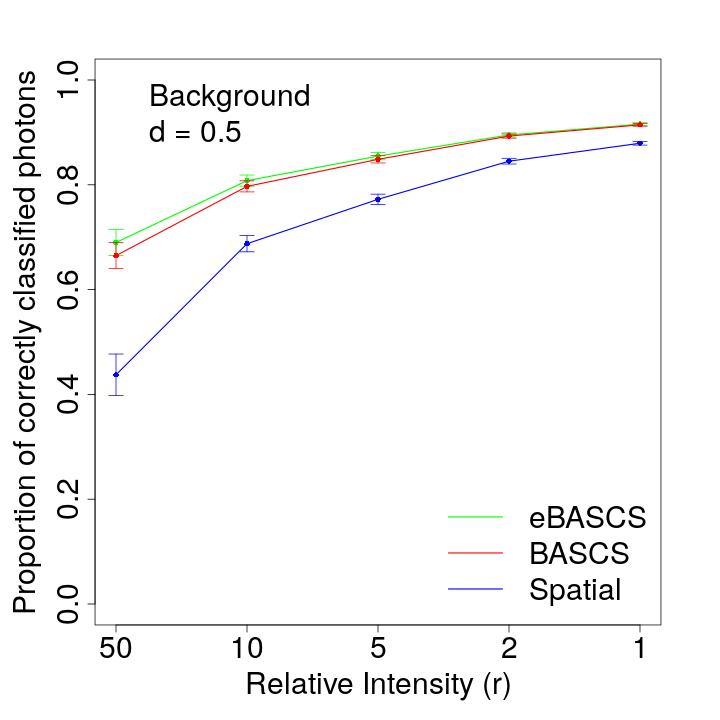}
   \end{minipage}
   \caption{Allocation recovery for the bright source (left), faint source (middle) and background (right) by \ebascs{} (green), \bascs{} (red) and \spatial{}  (blue), for simulation settings $(d = 0.5, r \in \{1,2,5,10,50\})$, averaged over the replicate data sets.
   }
   \label{tab:appperfeval0.5}
\end{figure}

\begin{figure}
   \begin{minipage}{0.3\linewidth}
     \centering
     \includegraphics[width=1\linewidth]{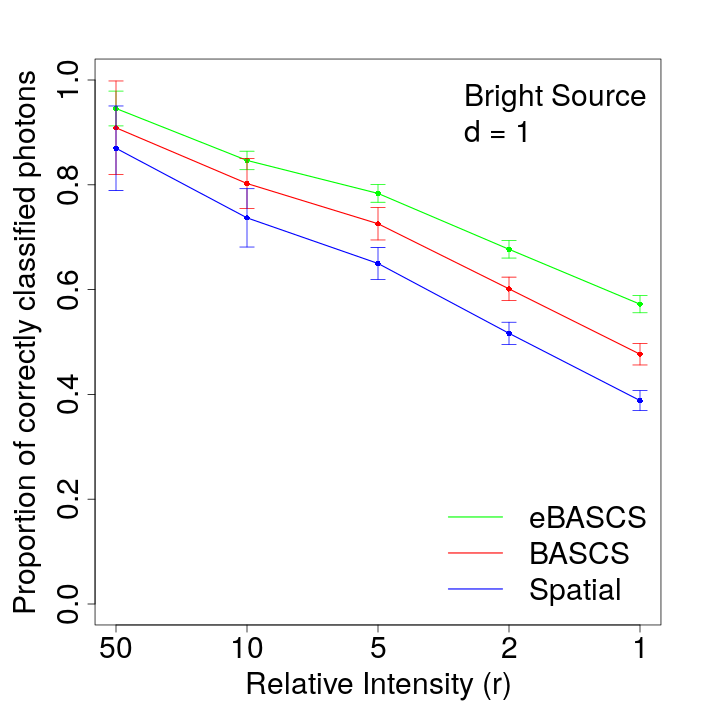}
   \end{minipage}\hfill
   \begin{minipage}{0.3\linewidth}
     \centering
     \includegraphics[width=1\linewidth]{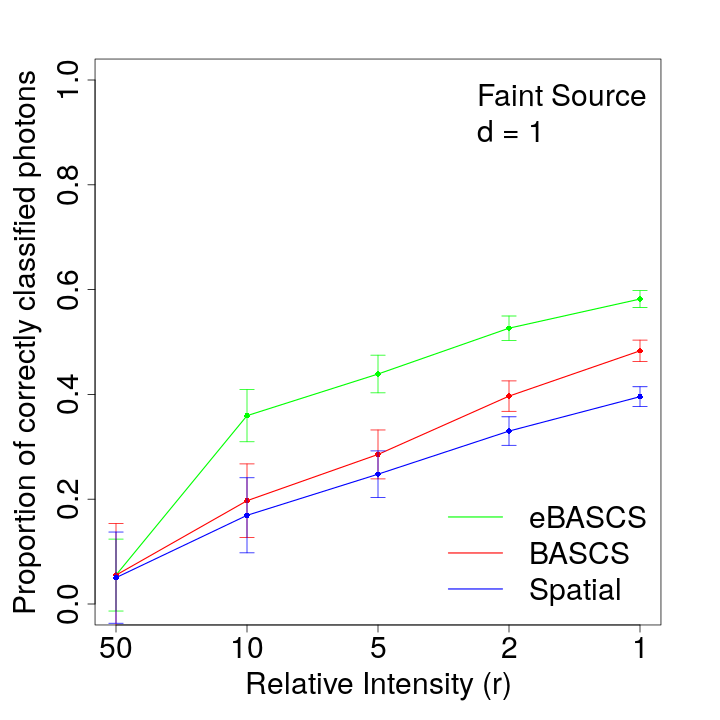}
   \end{minipage}\hfill
      \begin{minipage}{0.3\linewidth}
     \centering
     \includegraphics[width=1\linewidth]{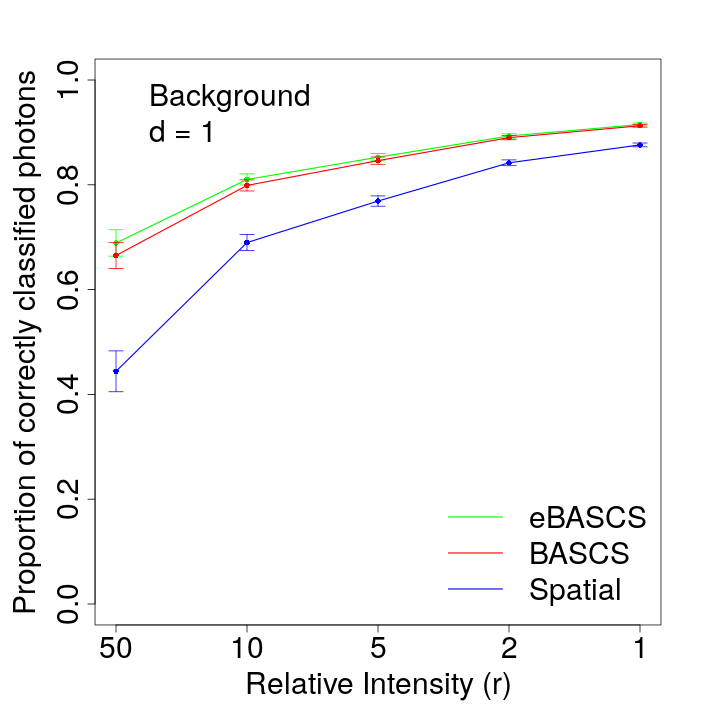}
   \end{minipage}
   \caption{As in Figure \ref{tab:appperfeval0.5}, for $d=1$.
   }
   \label{tab:appperfeval1}
\end{figure}

\begin{figure}
   \begin{minipage}{0.3\linewidth}
     \centering
     \includegraphics[width=1\linewidth]{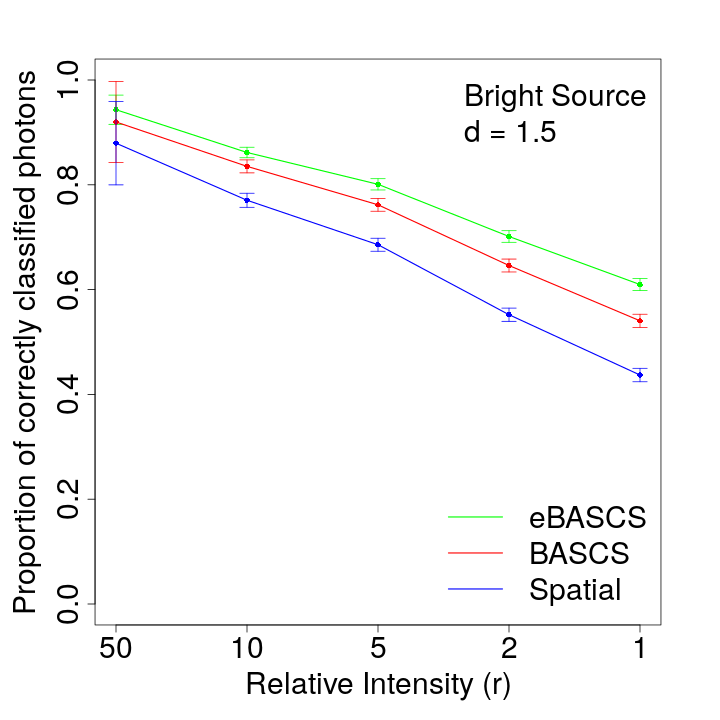}
   \end{minipage}\hfill
   \begin{minipage}{0.3\linewidth}
     \centering
     \includegraphics[width=1\linewidth]{plots/2papreppc_s2d1.5.png}
   \end{minipage}\hfill
      \begin{minipage}{0.3\linewidth}
     \centering
     \includegraphics[width=1\linewidth]{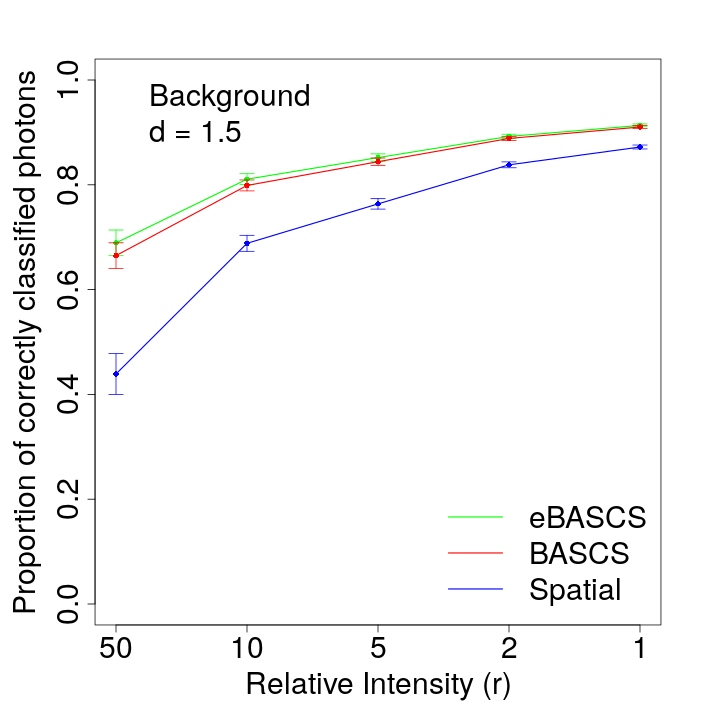}
   \end{minipage}
   \caption{As in Figure \ref{tab:appperfeval0.5}, for $d=1.5$.
   }
   \label{tab:appperfeval1.5}
\end{figure}

\begin{figure}
   \begin{minipage}{0.3\linewidth}
     \centering
     \includegraphics[width=1\linewidth]{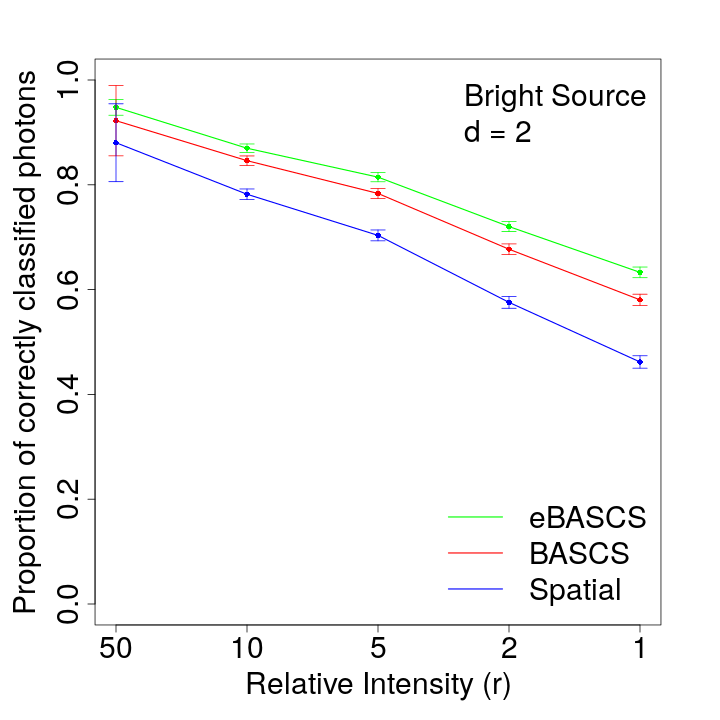}
   \end{minipage}\hfill
   \begin{minipage}{0.3\linewidth}
     \centering
     \includegraphics[width=1\linewidth]{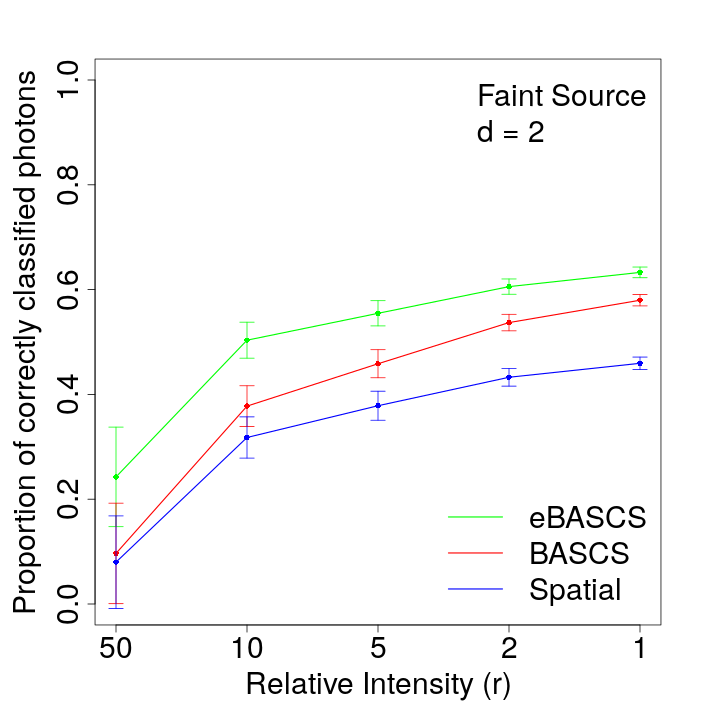}
   \end{minipage}\hfill
      \begin{minipage}{0.3\linewidth}
     \centering
     \includegraphics[width=1\linewidth]{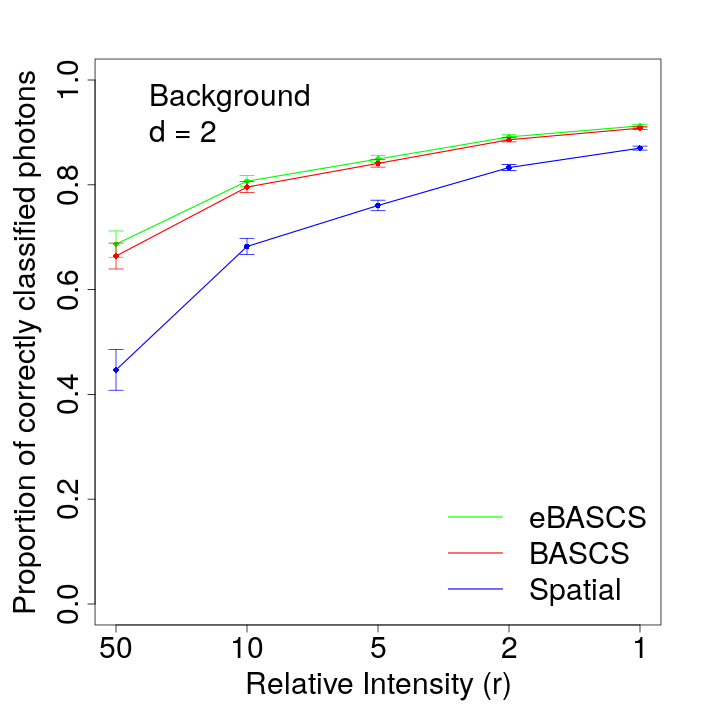}
   \end{minipage}
   \caption{As in Figure \ref{tab:appperfeval0.5}, for $d=2$.
   }
   \label{tab:appperfeval2}
\end{figure}

 Figures \ref{tab:appperfeval0.5}, \ref{tab:appperfeval1}, \ref{tab:appperfeval1.5} and \ref{tab:appperfeval2} present a graphical summary of the results given in Sections~\ref{sec:appagivens} and \ref{sec:appsgivena}. 
 
\clearpage

\section{Simulation~I Results}
\subsection{Simulation~I Allocation Recovery 
(fraction of events from a component that are correctly allocated to the same component)}

\label{sec:appagivens}
Tables \ref{tab:a1s1eB}, \ref{tab:a1s2eB}, \ref{tab:a1s3eB} give the allocation recovery of \ebascs{} for the bright source, faint source and background respectively.
Tables \ref{tab:a1s2rB}, \ref{tab:a1s2rB}, \ref{tab:a1s3rB} give the allocation recovery of \bascs{} for the bright source, faint source and background respectively.
Tables \ref{tab:a1s1sp}, \ref{tab:a1s2sB}, \ref{tab:a1s3sB} give the allocation recovery of \spatial{} for the bright source, faint source and background respectively.

\begin{table}
\parbox{.45\linewidth}{
\centering
\begin{tabular}{llllll}
\hline
  & $r$=50 & $r$=10 & $r$=5 & $r$=2 & $r$=1\\
\hline
  $d$=0.5 & 0.948 & 0.835 & 0.735 & 0.630 & 0.542\\
$d$=1 & 0.946 & 0.846 & 0.784 & 0.677 & 0.572\\
  $d$=1.5 & 0.943 & 0.862 & 0.801 & 0.701 & 0.610\\
$d$=2 & 0.947 & 0.870 & 0.816 & 0.720 & 0.633\\
\hline
\end{tabular}
\caption{Allocation recovery for the \textbf{bright source} by \ebascs{}, averaged over MCMC iterates and replicate data sets.
}
\label{tab:a1s1eB}}
\hfill
\parbox{.45\linewidth}{
\centering
\begin{tabular}{llllll}
\hline
  & $r$=50 & $r$=10 & $r$=5 & $r$=2 & $r$=1\\
\hline
  $d$=0.5 & 0.913 & 0.826 & 0.719 & 0.547 & 0.420\\
$d$=1 & 0.909 & 0.802 & 0.726 & 0.601 & 0.477\\
  $d$=1.5 & 0.920 & 0.835 & 0.762 & 0.646 & 0.540\\
$d$=2 & 0.922 & 0.846 & 0.783 & 0.677 & 0.580\\
\hline
\end{tabular}
\caption{As in Table \ref{tab:a1s1eB}, for \bascs{}.
}
\label{tab:a1s1rb}}
\end{table}
\begin{table}
\parbox{.45\linewidth}{
\centering
\begin{tabular}{llllll}
\hline
  & $r$=50 & $r$=10 & $r$=5 & $r$=2 & $r$=1\\
\hline
  $d$=0.5 & 0.025 & 0.201 & 0.383 & 0.479 & 0.534\\
$d$=1 & 0.055 & 0.359 & 0.439 & 0.526 & 0.582\\
  $d$=1.5 & 0.143 & 0.436 & 0.502 & 0.569 & 0.610\\
$d$=2 & 0.246 & 0.503 & 0.555 & 0.606 & 0.633\\
\hline
\end{tabular}
\caption{As in Table \ref{tab:a1s1eB}, for the \textbf{faint source}.
}
\label{tab:a1s2eB}
}
\hfill
\parbox{.45\linewidth}{
\centering
\begin{tabular}{llllll}
\hline
  & $r$=50 & $r$=10 & $r$=5 & $r$=2 & $r$=1\\
\hline
  $d$=0.5 & 0.045 & 0.070 & 0.141 & 0.287 & 0.380\\
$d$=1 & 0.055 & 0.197 & 0.285 & 0.397 & 0.483\\
  $d$=1.5 & 0.060 & 0.286 & 0.378 & 0.477 & 0.541\\
$d$=2 & 0.097 & 0.378 & 0.456 & 0.537 & 0.580\\
\hline
\end{tabular}
    \caption{As in Table \ref{tab:a1s1eB}, for the \textbf{faint source} by \bascs.
}
\label{tab:a1s2rB}
}
\end{table}
\begin{table}
\parbox{.45\linewidth}{
\centering
\begin{tabular}{llllll}
\hline
  & $r$=50 & $r$=10 & $r$=5 & $r$=2 & $r$=1\\
\hline
  $d$=0.5 & 0.690 & 0.808 & 0.855 & 0.895 & 0.916\\
$d$=1 & 0.689 & 0.810 & 0.853 & 0.893 & 0.915\\
  $d$=1.5 & 0.689 & 0.811 & 0.852 & 0.892 & 0.913\\
$d$=2 & 0.686 & 0.807 & 0.849 & 0.891 & 0.913\\
\hline
\end{tabular}
\caption{As in Table \ref{tab:a1s1eB}, for the \textbf{background}.
}
\label{tab:a1s3eB}
}
\hfill
\parbox{.45\linewidth}{
\centering
\begin{tabular}{llllll}
\hline
  & $r$=50 & $r$=10 & $r$=5 & $r$=2 & $r$=1\\
\hline
  $d$=0.5 & 0.665 & 0.797 & 0.849 & 0.893 & 0.914\\
$d$=1 & 0.665 & 0.799 & 0.846 & 0.890 & 0.913\\
  $d$=1.5 & 0.665 & 0.799 & 0.844 & 0.888 & 0.910\\
$d$=2 & 0.664 & 0.796 & 0.841 & 0.886 & 0.908\\
\hline
\end{tabular}
\caption{As in Table \ref{tab:a1s1eB}, for the \textbf{background} by \bascs{}.
}
\label{tab:a1s3rB}
}
\end{table}
\begin{table}
\parbox{.45\linewidth}{
\centering
\begin{tabular}{llllll}
\hline
  & $r$=50 & $r$=10 & $r$=5 & $r$=2 & $r$=1\\
\hline
  $d$=0.5 & 0.874 & 0.767 & 0.652 & 0.478 & 0.354\\
$d$=1 & 0.870 & 0.737 & 0.650 & 0.517 & 0.388\\
  $d$=1.5 & 0.879 & 0.770 & 0.686 & 0.552 & 0.437\\
$d$=2 & 0.880 & 0.782 & 0.703 & 0.576 & 0.462\\
\hline
\end{tabular}
\caption{As in Table \ref{tab:a1s1eB}, for \spatial{}.
}
\label{tab:a1s1sp}
}
\hfill
\parbox{.45\linewidth}{
\centering
\begin{tabular}{llllll}
\hline
  & $r$=50 & $r$=10 & $r$=5 & $r$=2 & $r$=1\\
\hline
  $d$=0.5 & 0.044 & 0.052 & 0.113 & 0.226 & 0.300\\
$d$=1 & 0.050 & 0.170 & 0.248 & 0.330 & 0.396\\
  $d$=1.5 & 0.050 & 0.247 & 0.319 & 0.392 & 0.437\\
$d$=2 & 0.080 & 0.318 & 0.378 & 0.433 & 0.459\\
\hline
\end{tabular}
\caption{As in Table \ref{tab:a1s1eB}, for the \textbf{faint source} by \spatial.
}
\label{tab:a1s2sB}
}
\end{table}
\begin{table}
\centering
\begin{tabular}{llllll}
\hline
  & $r$=50 & $r$=10 & $r$=5 & $r$=2 & $r$=1\\
\hline
  $d$=0.5 & 0.437 & 0.688 & 0.772 & 0.845 & 0.879\\
$d$=1 & 0.444 & 0.690 & 0.769 & 0.842 & 0.876\\
  $d$=1.5 & 0.439 & 0.688 & 0.764 & 0.838 & 0.872\\
$d$=2 & 0.447 & 0.682 & 0.761 & 0.833 & 0.870\\
\hline
\end{tabular}
\caption{As in Table \ref{tab:a1s1eB}, for the \textbf{background} by \spatial.
}
\label{tab:a1s3sB}
\end{table}

\clearpage
\subsection{Simulation~I Allocation Accuracy 
(fraction of events correctly allocated to the component)}
\label{sec:appsgivena}

Tables \ref{tab:allaccs1eB}, \ref{tab:allaccs2eB}, \ref{tab:allaccs3eB} give the allocation accuracy of \ebascs{} for the bright source, faint source and background respectively.
Tables \ref{tab:allaccs1rB}, \ref{tab:allaccs2rB}, \ref{tab:allaccs3rB} give the allocation accuracy of \bascs{} for the bright source, faint source and background respectively.
Tables \ref{tab:allaccs1sB}, \ref{tab:allaccs2sB}, \ref{tab:allaccs3sB} give the allocation accuracy of \spatial{} for the bright source, faint source and background respectively.





\begin{table}
\parbox{.45\linewidth}{
\centering
\begin{tabular}{llllll}
\hline
  & $r$=50 & $r$=10 & $r$=5 & $r$=2 & $r$=1\\
  \hline
  $d$=0.5 & 0.950 & 0.844 & 0.789 & 0.669 & 0.564\\
$d$=1 & 0.951 & 0.866 & 0.897 & 0.701 & 0.610\\
  $d$=1.5 & 0.954 & 0.876 & 0.822 & 0.725 & 0.645\\
$d$=2 & 0.956 & 0.883 & 0.835 & 0.756 & 0.670\\
\hline
\end{tabular}
\caption{Allocation accuracy for the \textbf{bright source} by \ebascs{}, averaged over MCMC iterates and replicate data sets.
}
\label{tab:allaccs1eB}
}
\hfill
\parbox{.45\linewidth}{
\centering
\begin{tabular}{llllll}
\hline
  & $r$=50 & $r$=10 & $r$=5 & $r$=2 & $r$=1\\
\hline
  $d$=0.5 & 0.946 & 0.821 & 0.725 & 0.562 & 0.428\\
$d$=1 & 0.947 & 0.835 & 0.753 & 0.620 & 0.504\\
  $d$=1.5 & 0.948 & 0.848 & 0.782 & 0.667 & 0.570\\
$d$=2 & 0.948 & 0.861 & 0.800 & 0.700 & 0.611\\
\hline
\end{tabular}
\caption{As in Table \ref{tab:allaccs1eB}, for \bascs.
}
\label{tab:allaccs1rB}
}
\end{table}
\begin{table}
\parbox{.45\linewidth}{
\centering
\begin{tabular}{llllll}
\hline
  & $r$=50 & $r$=10 & $r$=5 & $r$=2 & $r$=1\\
\hline
  $d$=0.5 & 0.0009 & 0.168 & 0.338 & 0.487 & 0.571\\
$d$=1 & 0.016 & 0.363 & 0.456 & 0.550 & 0.608\\
  $d$=1.5 & 0.078 & 0.459 & 0.535 & 0.603 & 0.641\\
$d$=2 & 0.186 & 0.527 & 0.579 & 0.647 & 0.672\\
\hline
\end{tabular}
\caption{As in Table \ref{tab:allaccs1eB}, for the \textbf{faint source}.
}
\label{tab:allaccs2eB}
}
\hfill
\parbox{.45\linewidth}{
\centering
\begin{tabular}{llllll}
\hline
  & $r$=50 & $r$=10 & $r$=5 & $r$=2 & $r$=1\\
\hline
  $d$=0.5 & 0.010 & 0.031 & 0.109 & 0.301 & 0.422\\
$d$=1 & 0.011 & 0.176 & 0.278 & 0.420 & 0.505\\
  $d$=1.5 & 0.026 & 0.311 & 0.403 & 0.510 & 0.566\\
$d$=2 & 0.063 & 0.394 & 0.479 & 0.565 & 0.614\\
\hline
\end{tabular}
\caption{As in Table \ref{tab:allaccs1eB}, for the \textbf{faint source} by \bascs.
}
\label{tab:allaccs2rB}
}
\end{table}

\begin{table}
\parbox{.45\linewidth}{
\centering
\begin{tabular}{llllll}
\hline
  & $r$=50 & $r$=10 & $r$=5 & $r$=2 & $r$=1\\
\hline
  $d$=0.5 & 0.632 & 0.782 & 0.828 & 0.877 & 0.900\\
$d$=1 & 0.635 & 0.781 & 0.828 & 0.876 & 0.899\\
  $d$=1.5 & 0.641 & 0.781 & 0.826 & 0.871 & 0.897\\
$d$=2 & 0.639 & 0.777 & 0.824 & 0.870 & 0.894\\
\hline
\end{tabular}
\caption{As in Table \ref{tab:allaccs1eB}, for the \textbf{background}.
}
\label{tab:allaccs3eB}
}
\hfill
\parbox{.45\linewidth}{
\centering
\begin{tabular}{llllll}
\hline
  & $r$=50 & $r$=10 & $r$=5 & $r$=2 & $r$=1\\
\hline
  $d$=0.5 & 0.611 & 0.770 & 0.822 & 0.875 & 0.900\\
$d$=1 & 0.612 & 0.770 & 0.824 & 0.874 & 0.897\\
  $d$=1.5 & 0.618 & 0.771 & 0.818 & 0.868 & 0.894\\
$d$=2 & 0.610 & 0.766 & 0.818 & 0.866 & 0.890\\
\hline
\end{tabular}
\caption{As in Table \ref{tab:allaccs1eB}, for the \textbf{background} by \bascs.
}
\label{tab:allaccs3rB}
}
\end{table}
\begin{table}
\parbox{.45\linewidth}{
\centering
\begin{tabular}{llllll}
\hline
  & $r$=50 & $r$=10 & $r$=5 & $r$=2 & $r$=1\\
\hline
  $d$=0.5 & 0.922 & 0.770 & 0.665 & 0.492 & 0.358\\
$d$=1 & 0.923 & 0.783 & 0.693 & 0.549 & 0.430\\
  $d$=1.5 & 0.923 & 0.797 & 0.719 & 0.594 & 0.489\\
$d$=2 & 0.927 & 0.811 & 0.739 & 0.618 & 0.515\\
\hline
\end{tabular}
\caption{As in Table \ref{tab:allaccs1eB}, by \spatial.
}
\label{tab:allaccs1sB}
}
\hfill
\parbox{.45\linewidth}{
\centering
\begin{tabular}{llllll}
\hline
  & $r$=50 & $r$=10 & $r$=5 & $r$=2 & $r$=1\\
\hline
  $d$=0.5 & 0.007 & 0.013 & 0.102 & 0.236 & 0.352\\
$d$=1 & 0.005 & 0.142 & 0.251 & 0.368 & 0.434\\
  $d$=1.5 & 0.014 & 0.268 & 0.355 & 0.436 & 0.481\\
$d$=2 & 0.041 & 0.357 & 0.424 & 0.484 & 0.516\\
\hline
\end{tabular}
\caption{As in Table \ref{tab:allaccs1eB}, for the \textbf{faint source} by \spatial.
}
\label{tab:allaccs2sB}
}
\end{table}
\begin{table}
\centering
\begin{tabular}{llllll}
\hline
  & $r$=50 & $r$=10 & $r$=5 & $r$=2 & $r$=1\\
\hline
  $d$=0.5 & 0.347 & 0.637 & 0.730 & 0.815 & 0.853\\
$d$=1 & 0.347 & 0.635 & 0.726 & 0.809 & 0.849\\
  $d$=1.5 & 0.344 & 0.634 & 0.719 & 0.802 & 0.843\\
$d$=2 & 0.352 & 0.627 & 0.717 & 0.796 & 0.836\\
\hline
\end{tabular}
\caption{As in Table \ref{tab:allaccs1eB}, for the \textbf{background} by \spatial.
}
\label{tab:allaccs3sB}
\end{table}

\clearpage

\section{Simulation~I: Sensitivity of posterior source location to parameter settings}
\label{sec:appsensloc}
\begin{figure}
    \centering
    \includegraphics[width=0.21\linewidth, height = 0.08\textheight]{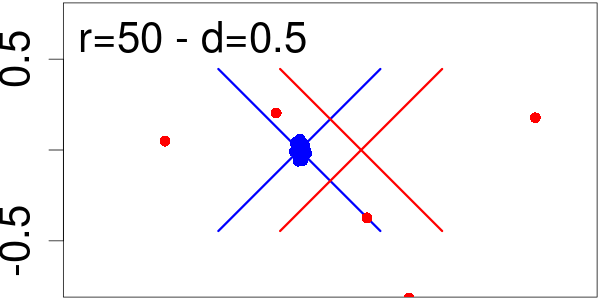}\hspace{0.2mm}
    \vspace{0.1mm}
    \includegraphics[width=0.21\linewidth, height = 0.08\textheight]{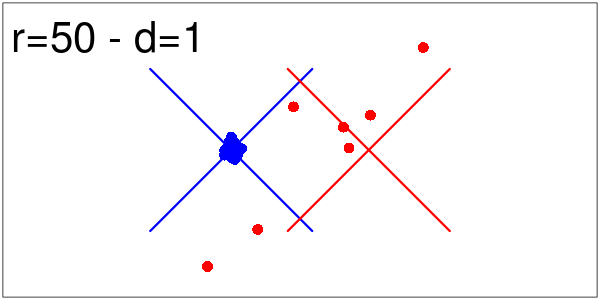}\hspace{0.2mm}
    \includegraphics[width=0.21\linewidth, height = 0.08\textheight]{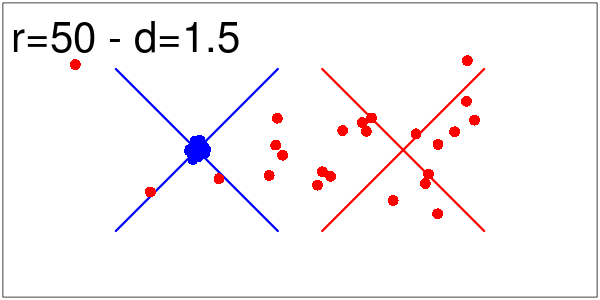}\hspace{0.2mm}
    \includegraphics[width=0.21\linewidth, height = 0.08\textheight]{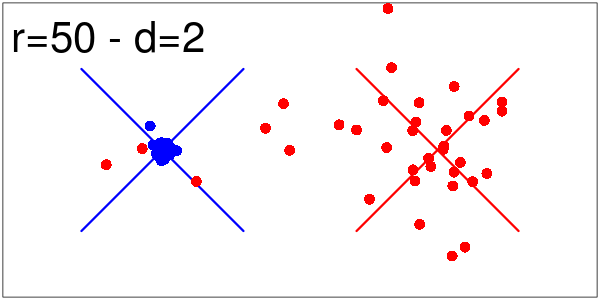} \vspace{0.1mm}
    \includegraphics[width=0.21\linewidth, height = 0.08\textheight]{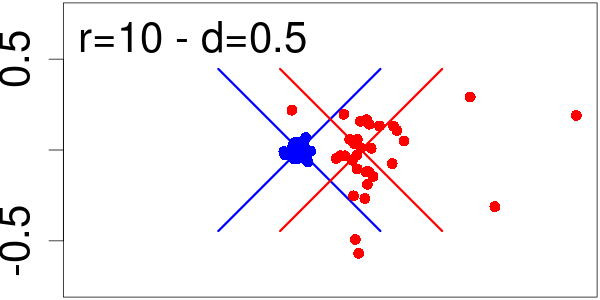}\hspace{0.2mm}
    \includegraphics[width=0.21\linewidth, height = 0.08\textheight]{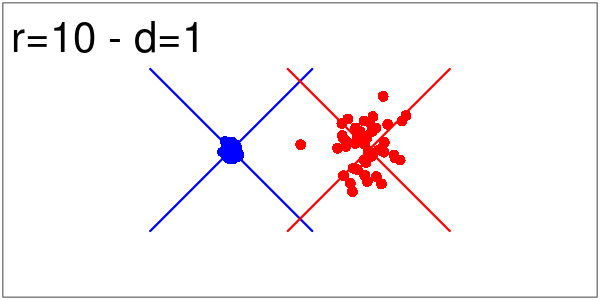}\hspace{0.2mm}
    \includegraphics[width=0.21\linewidth, height = 0.08\textheight]{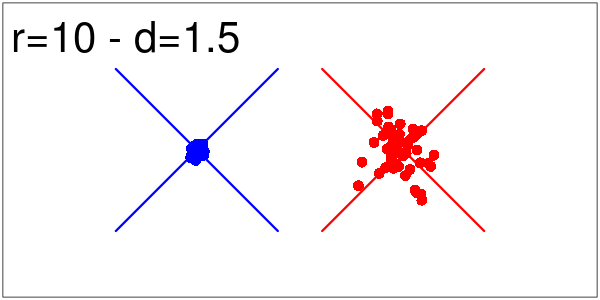}\hspace{0.2mm}
    \includegraphics[width=0.21\linewidth, height = 0.08\textheight]{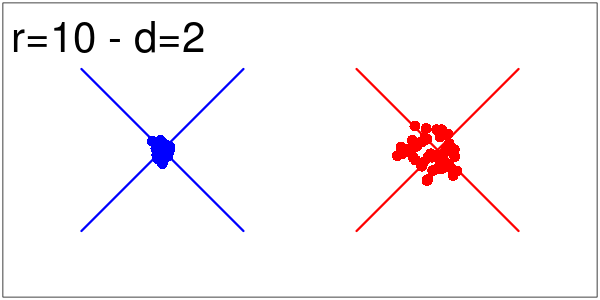}\vspace{0.1mm}
    \includegraphics[width=0.21\linewidth, height = 0.08\textheight]{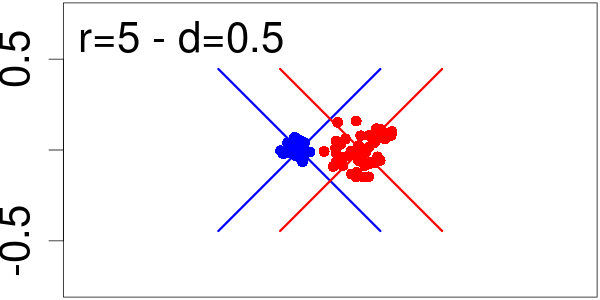}\hspace{0.2mm}
    \includegraphics[width=0.21\linewidth, height = 0.08\textheight]{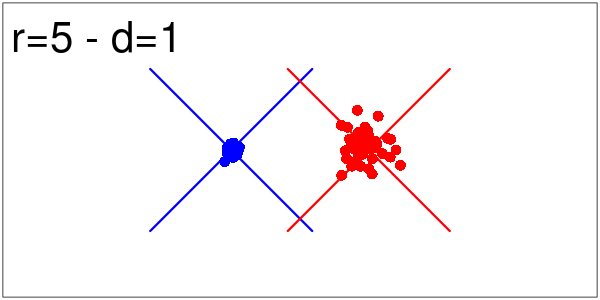}\hspace{0.2mm}
    \includegraphics[width=0.21\linewidth, height = 0.08\textheight]{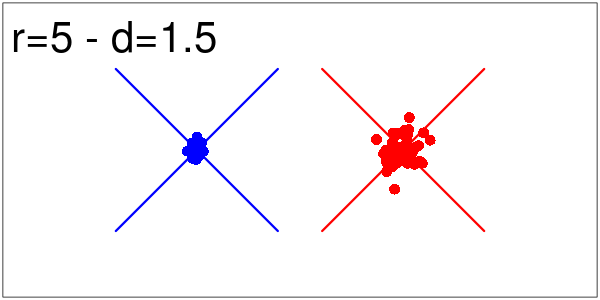}\hspace{0.2mm}
    \includegraphics[width=0.21\linewidth, height = 0.08\textheight]{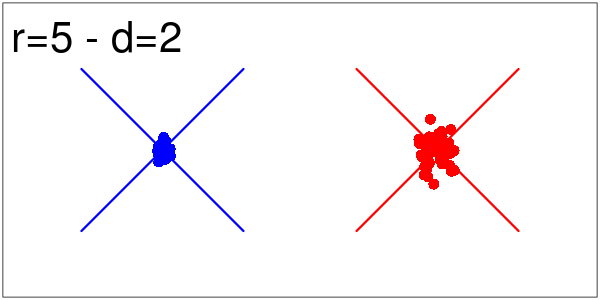}\vspace{0.1mm}
    \includegraphics[width=0.21\linewidth, height = 0.08\textheight]{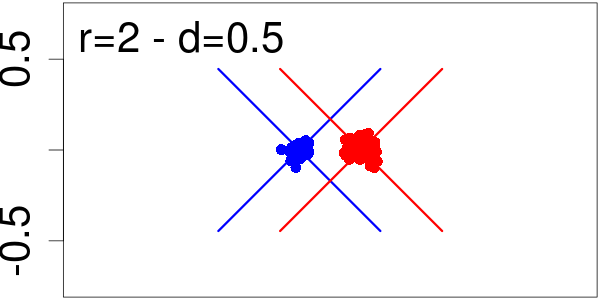}\hspace{0.2mm}
    \includegraphics[width=0.21\linewidth, height = 0.08\textheight]{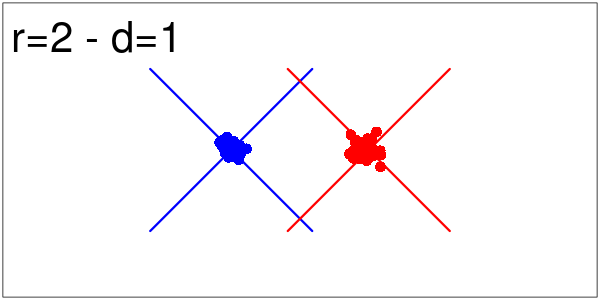}\hspace{0.2mm}
    \includegraphics[width=0.21\linewidth, height = 0.08\textheight]{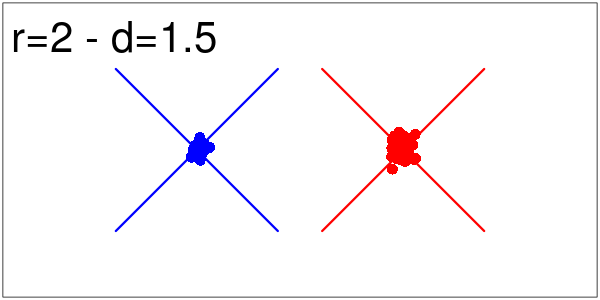}\hspace{0.2mm}
    \includegraphics[width=0.21\linewidth, height = 0.08\textheight]{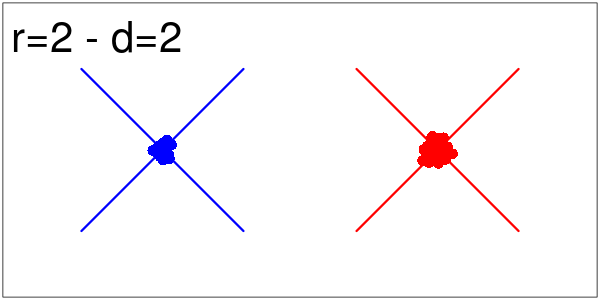}\hspace{0.3mm} \vspace{0.1mm}
    \hspace{0.1mm}
    \includegraphics[width=0.21\linewidth, height = 0.09\textheight]{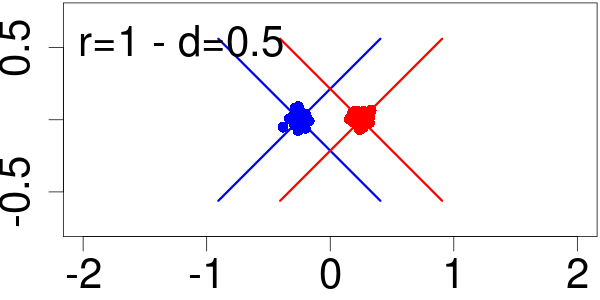}\hspace{0.2mm}
    \includegraphics[width=0.21\linewidth, height = 0.09\textheight]{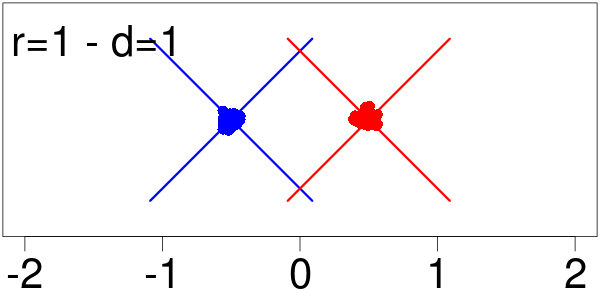}\hspace{0.2mm}
    \includegraphics[width=0.21\linewidth, height = 0.09\textheight]{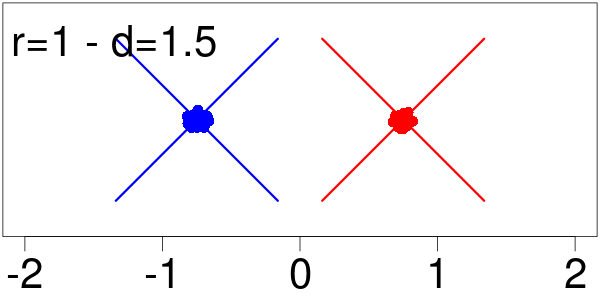}\hspace{0.2mm}
    \includegraphics[width=0.21\linewidth, height = 0.09\textheight]{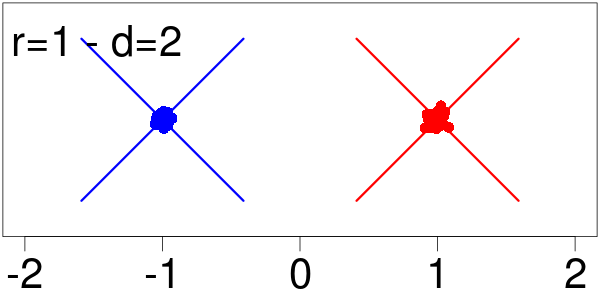}
    \caption{Sensitivity of location determination under \ebascs{} as a function of source separation ($d$) and relative strength ($r$). Red and blue dots give the mean posterior locations for each dataset replicate of the bright and faint sources, respectively, under the simulation setting indicated in the top-left corner of the box. The large “X”s of corresponding color indicate the true locations.}
    \label{tab:sensloc}
    \includegraphics[width=0.21\linewidth, height = 0.08\textheight]{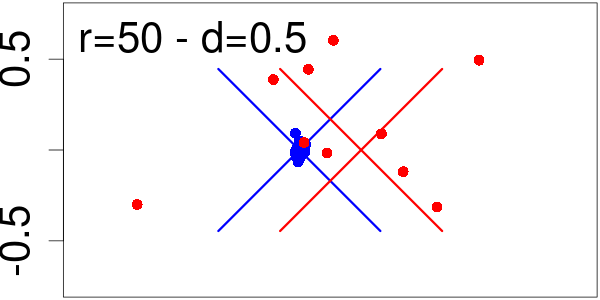}\hspace{0.2mm} \vspace{0.1mm}
    \includegraphics[width=0.21\linewidth, height = 0.08\textheight]{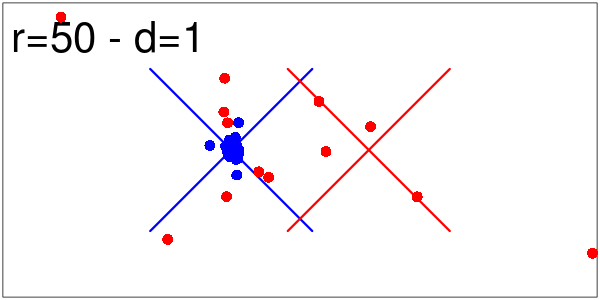}\hspace{0.2mm}
    \includegraphics[width=0.21\linewidth, height = 0.08\textheight]{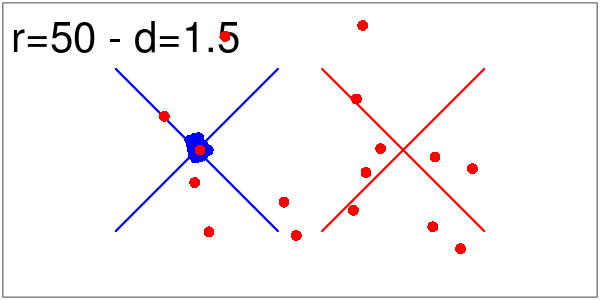}\hspace{0.2mm}
    \includegraphics[width=0.21\linewidth, height = 0.08\textheight]{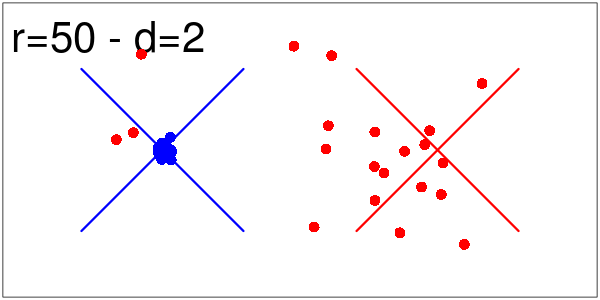} \vspace{0.1mm}
    \includegraphics[width=0.21\linewidth, height = 0.08\textheight]{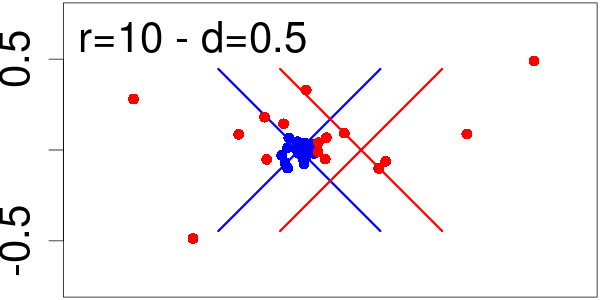}\hspace{0.2mm}
    \includegraphics[width=0.21\linewidth, height = 0.08\textheight]{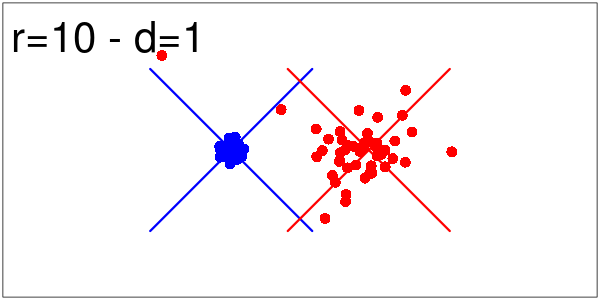}\hspace{0.2mm}
    \includegraphics[width=0.21\linewidth, height = 0.08\textheight]{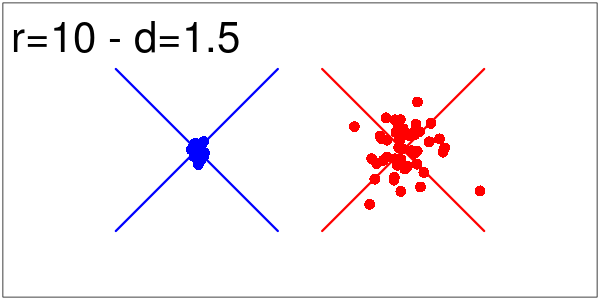}\hspace{0.2mm}
    \includegraphics[width=0.21\linewidth, height = 0.08\textheight]{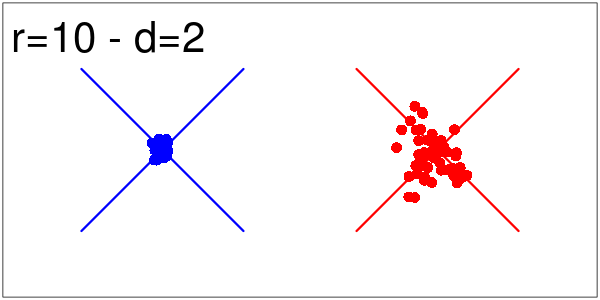}
    \includegraphics[width=0.21\linewidth, height = 0.08\textheight]{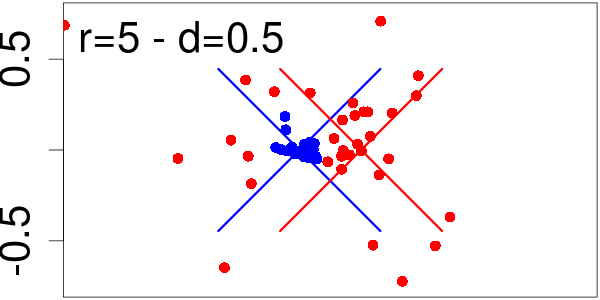}\hspace{0.2mm}
    \includegraphics[width=0.21\linewidth, height = 0.08\textheight]{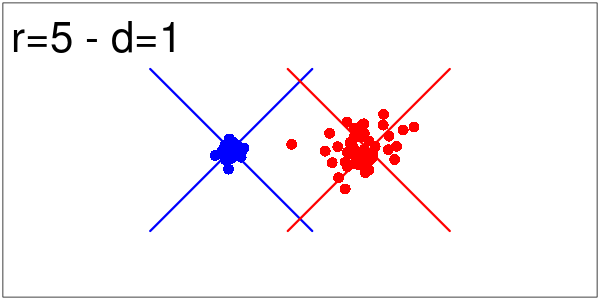}\hspace{0.2mm}
    \includegraphics[width=0.21\linewidth, height = 0.08\textheight]{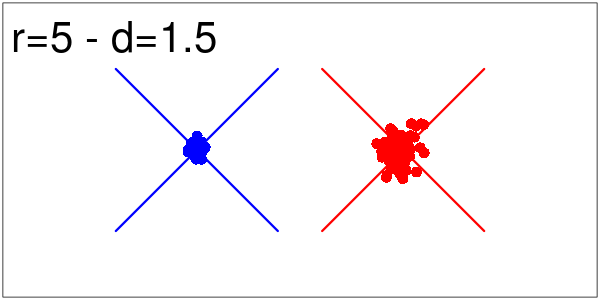}\hspace{0.2mm}
    \includegraphics[width=0.21\linewidth, height = 0.08\textheight]{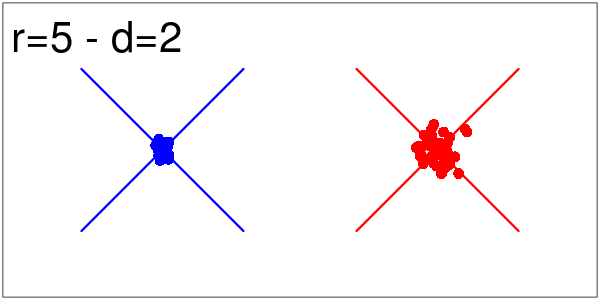}\vspace{0.1mm}
    \includegraphics[width=0.21\linewidth, height = 0.08\textheight]{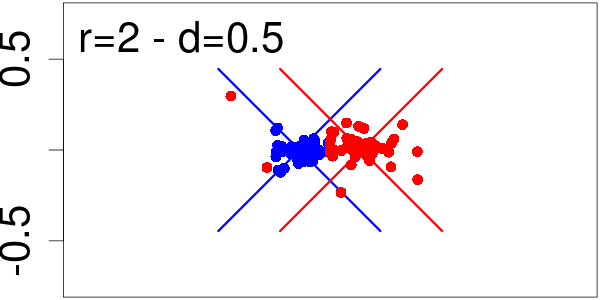}\hspace{0.2mm}
    \includegraphics[width=0.21\linewidth, height = 0.08\textheight]{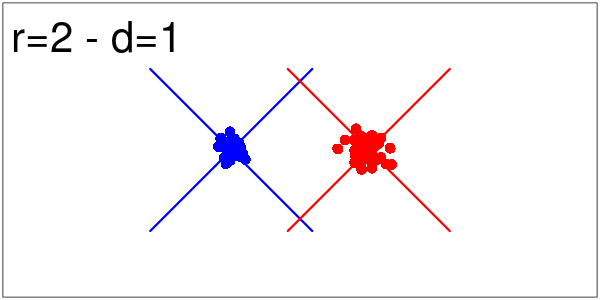}\hspace{0.2mm}
    \includegraphics[width=0.21\linewidth, height = 0.08\textheight]{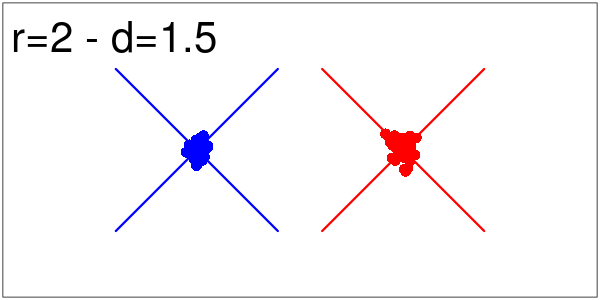}\hspace{0.2mm}
    \includegraphics[width=0.21\linewidth, height = 0.08\textheight]{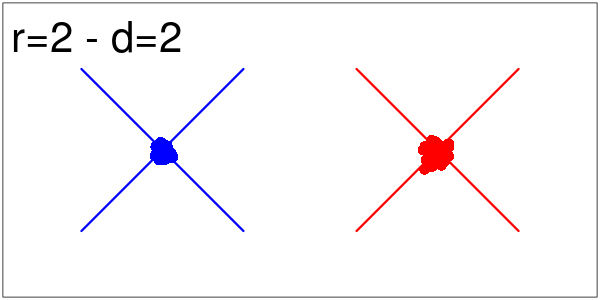}\hspace{0.3mm} \vspace{0.1mm}
    \hspace{0.1mm}
    \includegraphics[width=0.21\linewidth, height = 0.09\textheight]{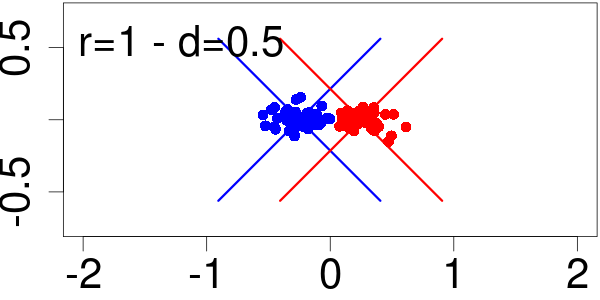}\hspace{0.2mm}
    \includegraphics[width=0.21\linewidth, height = 0.09\textheight]{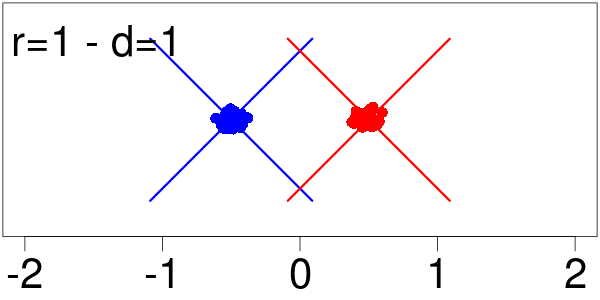}\hspace{0.2mm}
    \includegraphics[width=0.21\linewidth, height = 0.09\textheight]{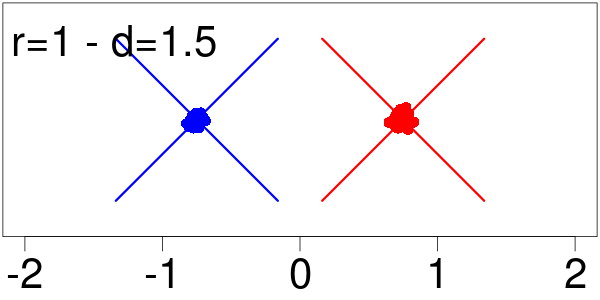}\hspace{0.2mm}
    \includegraphics[width=0.21\linewidth, height = 0.09\textheight]{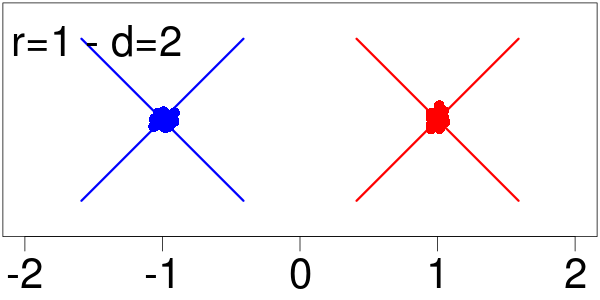}
    \caption{As in Figure \ref{tab:sensloc}, for \bascs.
    }
    \label{tab:sensloc2}
\end{figure}

Figures \ref{tab:sensloc} and \ref{tab:sensloc2} show the true location of sources and their mean posterior locations under \ebascs{} (top part) and \bascs{} (bottom part) for all data set replicates under the simulation parameter setting indicated in the top left corner of the plot (for Simulation~I). Each dot represents the mean posterior location of the source (blue for bright source, red for faint source) for  one dataset replicate, and the large "X"'s of corresponding color indicate the true locations.

\clearpage

\section{Simulation~I: Parameter Estimates}
\label{sec:appsimulationpara}

Table \ref{tab:highsimpara} shows the true and estimated parameter values for a single dataset replicate under the simulation settings $d= 0.5, r = 1$, for the Simulation~I. The posterior means of the fitted location and intensity parameters are closer to their respective true values when inferred by \ebascs{} compared to \bascs{} and \spatial{}. The posterior quantile intervals are also narrower for \ebascs{}. This shows that \ebascs{} is able to estimate the model parameters shared with \bascs{} and \spatial{} more accurately and more confidently. 


\begin{table*}
\small
\centering
\begin{tabular}{|l|l||ll||ll||ll|}
\hline
  & & \ebascs{} & & \bascs{} & & \spatial & \\
  & \textbf{Truth} & mean & (q16,q84) & mean & (q16,q84) & mean & (q16,q18)\\
\hline
  $x_1$ & -0.25 &-0.257 & (-0.30,-0.210) & -0.270 & (-0.358, -0.189) &-0.279 & (-0.365 , -0.192)\\
$y_1$ & 0 & 0.032 & (0.006,0.058) & 0.037 & (-0.008,0.081) &  0.046 & (-0.005,0.097)\\
  $x_2$ & 0.25 & 0.229 & (0.198,0.258) & 0.208 & (0.152,0.271) & 0.189 & (0.134,0.245)\\
$y_2$ & 0 & 0.025 & (0.002,0.047) & 0.019 & (-0.014,0.052) & 0.022 & (-0.010,0.054)\\
\hline
  $w_1$ & 0.138 & 0.121 & (0.105,0.136) & 0.113 & (0.079,0.146) & 0.099 & (0.068,0.130) \\
$w_2$ & 0.138 & 0.145 & (0.130,0.161) & 0.153 & (0.120,0.187) & 0.156 & (0.126,0.187) \\
  $w_3$ & 0.724 & 0.734 & (0.729,0.739) & 0.734 & (0.729,0.739) & 0.745 & (0.739,0.750)\\
  \hline
$\gamma_{1}$ & 1832 & 1778.132 & (1732.254,1825.486)& 1831.655 & (1821.378,1841.74) &* &*\\
  $\alpha_{1}$ & 3.18 & 3.212 & (3.017,3.407) & 3.313 & (3.166,3.519) &* &*\\
$\gamma_{2}$ & 1832 & 1871.892 & (1831.783,1911.981)& 1831.113 & (1817.616,1843.494) &* &*\\

  $\alpha_{2}$ & 3.18 & 3.336 & (3.158,3.512) & 3.208 & (3.121,3.323) &* &* \\
  \hline
$\lambda_{1,1}$ & 0.05 & 0.039 & (0.011,0.069) & *&* &*&*\\
  $\lambda_{1,2}$ & 0.15 & 0.199 & (0.170,0.227) & *&* &*&* \\
$\lambda_{1,3}$ & 0.3 & 0.291 & (0.262,0.318) & *&* &*&* \\
  $\lambda_{1,4}$ & 0.5 & 0.471 & (0.433,0.508) & *&* &*&* \\
\hline
$\lambda_{2,1}$ & 0.5 & 0.511 & (0.465,0.557) & *&* &*&* \\
  $\lambda_{2,2}$ & 0.3 & 0.234 & (0.210,0.258) & *&* &*&* \\
$\lambda_{2,3}$ & 0.15 & 0.160 & (0.134,0.186) & *&* &*&* \\
  $\lambda_{2,4}$ & 0.05 & 0.095 & (0.060,0.130) & *&* &*&* \\
  \hline
$\lambda_{3,1}$ & 0.25 & 0.245 & (0.240,0.249) & *&* &*&* \\
  $\lambda_{3,2}$ & 0.25 & 0.252 & (0.247,0.256) & *&* &*&* \\
$\lambda_{3,3}$ & 0.25 & 0.257 & (0.252,0.261) & *&* &*&* \\
  $\lambda_{3,4}$ & 0.25 & 0.247 & (0.242,0.252) & *&* &*&* \\
\hline
\end{tabular}
\caption{\ebascs{} (left), \bascs{} (middle) and \spatial{} (right) parameter estimates, resulting from an application of the models to a simulation dataset with settings $d=0.5, r = 1$. Parameters $x_j$ and $y_j$ denote the spatial coordinates of the sources, $w_j$ their relative intensities ($j=1$ corresponds to the bright source, $j=2$ to the faint source and $j=3$ to the background). $\gamma_j$ and $\alpha_j$ respectively denote the mean and shape parameters of the {\emph{gamma}} distributions. $\lambda_{j,k}$ denotes the relative time intensity of source $j$ in time bin $k$. The "truth" column gives the parameters values used to generate the data. The "mean" column gives the posterior mean of the corresponding parameters, and the "(q16,q84)" column reports the 16$\%$ and 84$\%$ posterior quantiles. 
}
\label{tab:highsimpara}
\end{table*}

\clearpage

\clearpage
\section{Simulation~II Results}
\label{sec:applowback}

\subsection{Simulation~II: Allocation Recovery}
\label{subsec:lowbackagivens}

Tables \ref{tab:recovsim2eB}, \ref{tab:recovsim2rB}, \ref{tab:recovsim2sp} and \ref{tab:recovsim2st} give the allocation recoveries of \ebascs{}, \bascs{}, \spatial{} and \spacetime{} (respectively) in Simulation~II.


\begin{table}
\parbox{.45\linewidth}{
\centering
\begin{tabular}{llllll}
\hline
  & $r$=50 & $r$=10 & $r$=5 & $r$=2 & $r$=1\\
\hline
  Bright Source & 0.955 & 0.917 & 0.883 & 0.826 & 0.763\\
Faint Source & 0.077 & 0.410 & 0.539 & 0.682 & 0.766\\
  Background & 0.616 & 0.609 & 0.588 & 0.560 & 0.525\\
\hline
\end{tabular}
\caption{Allocation recovery by \ebascs{}, averaged over MCMC iterates and replicate data sets.
}
\label{tab:recovsim2eB}
}
\hfill
\parbox{.45\linewidth}{
\centering
\begin{tabular}{llllll}
\hline
  & $r$=50 & $r$=10 & $r$=5 & $r$=2 & $r$=1\\
\hline
  Bright Source & 0.926 & 0.889 & 0.837 & 0.738 & 0.631\\
Faint Source & 0.057 & 0.197 & 0.331 & 0.509 & 0.638\\
  Background & 0.592 & 0.596 & 0.579 & 0.557 & 0.525\\
\hline
\end{tabular}
\caption{As in Table \ref{tab:recovsim2eB}, for \bascs{}.}
\label{tab:recovsim2rB}
}
\end{table}

\begin{table}
\parbox{.45\linewidth}{
\centering
\begin{tabular}{llllll}
\hline
  & $r$=50 & $r$=10 & $r$=5 & $r$=2 & $r$=1\\
\hline
  Bright Source & 0.892 & 0.862 & 0.812 & 0.719 & 0.617\\
Faint Source & 0.061 & 0.186 & 0.323 & 0.496 & 0.624\\
  Background & 0.335 & 0.337 & 0.313 & 0.289 & 0.251\\
\hline
\end{tabular}
\caption{As in Table \ref{tab:recovsim2eB}, for \spatial{}.}
\label{tab:recovsim2sp}
}
\hfill
\parbox{.45\linewidth}{
\centering
\begin{tabular}{llllll}
\hline
  & $r$=50 & $r$=10 & $r$=5 & $r$=2 & $r$=1\\
\hline
  Bright Source & 0.895 & 0.891 & 0.862 & 0.808 & 0.747\\
Faint Source & 0.148 & 0.404 & 0.524 & 0.666 & 0.751\\
  Background & 0.374 & 0.354 & 0.323 & 0.283 & 0.239\\
\hline
\end{tabular}
\caption{As in Table \ref{tab:recovsim2eB}, for \spacetime{}.}
\label{tab:recovsim2st}
}
\end{table}

\clearpage
\subsection{Simulation~II: Allocation Accuracy}
\label{subsec:lowbacksgivena}

Tables \ref{tab:accursim2eB}, \ref{tab:accursim2rB}, \ref{tab:accursim2sB} and \ref{tab:accursim2st} give the allocation accuracies of \ebascs{}, \bascs{}, \spatial{} and \spacetime{} (respectively) in Simulation~II.


\begin{table}
\parbox{.45\linewidth}{
\centering
\begin{tabular}{llllll}
\hline
  & $r$=50 & $r$=10 & $r$=5 & $r$=2 & $r$=1\\
\hline
  Bright Source & 0.962 & 0.927 & 0.896 & 0.837 & 0.769\\
Faint Source & 0.037 & 0.413 & 0.519 & 0.678 & 0.771\\
  Background & 0.564 & 0.551 & 0.519 & 0.503 & 0.478\\
\hline
\end{tabular}
\caption{Allocation accuracy of \ebascs{}, for $d=1$, averaged over MCMC iterates and replicate datasets.
}
\label{tab:accursim2eB}
}
\hfill
\parbox{.45\linewidth}{
\centering
\begin{tabular}{llllll}
\hline
  & $r$=50 & $r$=10 & $r$=5 & $r$=2 & $r$=1\\
\hline
  Bright Source & 0.960 & 0.903 & 0.850 & 0.747 & 0.638\\
Faint Source & 0.012 & 0.188 & 0.323 & 0.509 & 0.639\\
  Background & 0.549 & 0.526 & 0.527 & 0.499 & 0.474\\
\hline
\end{tabular}
\caption{As in Table \ref{tab:accursim2eB}, for \bascs{}.}
\label{tab:accursim2rB}
}
\end{table}
\begin{table}
\parbox{.45\linewidth}{
\centering
\begin{tabular}{llllll}
\hline
  & $r$=50 & $r$=10 & $r$=5 & $r$=2 & $r$=1\\
\hline
  Bright Source & 0.946 & 0.890 & 0.840 & 0.741 & 0.635\\
Faint Source & 0.016 & 0.183 & 0.326 & 0.510 & 0.636\\
  Background & 0.246 & 0.228 & 0.206 & 0.179 & 0.142\\
\hline
\end{tabular}
\caption{As in Table \ref{tab:accursim2eB}, for \spatial{}.}
\label{tab:accursim2sB}
}
\hfill
\parbox{.45\linewidth}{
\centering
\begin{tabular}{llllll}
\hline
  & $r$=50 & $r$=10 & $r$=5 & $r$=2 & $r$=1\\
\hline
  Bright Source & 0.950 & 0.917 & 0.890 & 0.830 & 0.765\\
Faint Source & 0.050 & 0.410 & 0.513 & 0.673 & 0.764\\
  Background & 0.284 & 0.252 & 0.236 & 0.187 & 0.144\\
\hline
\end{tabular}
\caption{As in Table \ref{tab:accursim2eB}, for \spacetime{}.}
\label{tab:accursim2st}
}
\end{table}

\clearpage

\section{\hbc\,A parameter estimates}
\label{sec:paramshbc515}


Table \ref{tab:params1} compares estimates for model parameters (fitted to the \hbc\,A data) shared by \ebascs{} and \bascs{} (i.e., spatial and spectral parameters).

\begin{table*}
    \centering
    \small
    \begin{tabular}{|l|ll|ll|}
    \hline
     &  \ebascs{} & & \bascs{} & \\
     & mean & (q16,q84) & mean & (q16,q84) \\
    \hline
    $x_1$ & 5.696 & (5.688, 5.705) & 5.697 & (5.688, 5.704)\\
    $y_1$ & 7.962 & (7.955, 7.969) & 7.961 & (7.954, 7.969)\\
    $x_2$ & 6.883 & (6.871, 6.894) & 6.881 & (6.871, 6.895)\\
    $y_2$ & 7.669 & (7.658, 7.679) & 7.670 & (7.659, 7.680)\\
    \hline
    $w_1$ & 0.614 & (0.607, 0.622) & 0.615 & (0.607, 0.622)\\
    $w_2$ & 0.386 & (0.378, 0.393) & 0.385 & (0.378, 0.393)\\
    $w_3$ & 0.000 & (0.000, 0.000) & 0.000 & (0.000, 0.000)\\
    \hline
    $\gamma_{1,1}$ & 1110.309 & (1092.170, 1128.832) & 1116.860 & (1094.428, 1134.366)\\
    $\gamma_{1,2}$ & 2278.711 & (2207.070, 2351.687) & 2328.049 & (2223.891, 2379.936)\\
    $\alpha_{1,1}$ & 10.440 & (9.675, 11.197) & 10.227 & (9.482, 11.042)\\
    $\alpha_{1,2}$ & 3.516 & (3.376, 3.655) & 3.631 & (3.396, 3.691)\\
    $\gamma_{2,1}$ & 1110.350 & (1084.721, 1135.518) & 1112.911 & (1077.943, 1133.171)\\
    $\gamma_{2,2}$ & 2244.203 & (2132.280, 2353.022) & 2236.035 & (2106.015, 2334.583)\\
    $\alpha_{2,1}$ & 11.643 & (10.339, 12.949) & 11.980 & (10.468, 13.490)\\
    $\alpha_{2,2}$ & 3.680 & (3.459, 3.892) & 3.673 & (3.447, 3.867)\\
    $\pi_1$ & 0.601 & (0.572, 0.630) & 0.612 & (0.578, 0.641)\\
    $\pi_2$ & 0.627 & (0.583, 0.671) & 0.619 & (0.571, 0.665)\\
    \hline
    \end{tabular}
    \caption{\hbc\,A fitted spatial and spectral parameters under \ebascs{} (left) and \bascs{} (right). $x_j$ and $y_j$ denote the spatial coordinates of source $j$, $w_j$ denotes the relative intensity. $\gamma_{j,l}$ denotes the mean parameter of the $l^{th}$ \emph{gamma} distribution component of source $j$'s spectral model (similarly with $\alpha_{j,l}$ for the shape parameter). $\pi_j$ denotes the mixture weight for source $j$'s mixture-of-\emph{gammas} spectral model. $j=1$ corresponds to \hbc\,Aa, $j=2$ to \hbc\,Ab and $j=3$ to the background. The "mean" columns give the posterior means of the corresponding parameters, and columns "(q16,q84)" give the 16$\%$ and 84 $\%$ posterior quantiles.}
    \label{tab:params1}
\end{table*}

\clearpage
\section{UV Ceti parameter estimates}
Table \ref{tab:uvspatial} gives estimates for the \spatial{} model parameters, when fitted to the \uvcet\ data.Table \ref{tab:SpaceTimeUVCet} gives estimates for the \spacetime{} model parameters.


\begin{table}
\centering
\begin{tabular}{|llll|}
\hline
  & mean & q16 & q84\\
\hline
  $x_1$ & 28.074 & 28.052 & 28.096\\
$y_1$ & 43.503 & 43.481 & 43.525\\
  $x_2$ & 39.008 & 38.975 & 39.042\\
$y_2$ & 48.199 & 48.164 & 48.235\\
\hline
  $w_1$ & 0.659 & 0.654 & 0.664\\
$w_2$ & 0.286 & 0.282 & 0.291\\
  $w_3$ & 0.055 & 0.052 & 0.058\\
\hline
\end{tabular}
\caption{\uvcet{} fitted parameters under the \spatial{} model. $x_j$ and $y_j$ denote the spatial coordinates of source $j$, $w_j$ denotes the relative intensity. $j=1$ corresponds to \uvcet\,B, $j=2$ corresponds to \uvcet\,A and $j=3$ corresponds to the background.The "mean" column gives the posterior means of the corresponding parameters, and the "(q16,q84)" column gives the 16$\%$ and 84 $\%$ posterior quantiles.}
\label{tab:uvspatial}
\end{table}
\begin{table}
\centering
\begin{tabular}{llll}
\hline
  & mean & q16 & q84\\
\hline
  $x_1$ & 28.074 & 28.052 & 28.095\\
$y_1$ & 43.503 & 43.481 & 43.526\\
  $x_2$ & 39.011 & 38.977 & 39.045\\
$y_2$ & 48.198 & 48.163 & 48.233\\
\hline
  $w_1$ & 0.660 & 0.655 & 0.665\\
$w_2$ & 0.286 & 0.282 & 0.291\\
  $w_3$ & 0.053 & 0.050 & 0.056\\
  \hline
$\lambda_{1,1}$ & 0.064 & 0.061 & 0.067\\
  $\lambda_{1,2}$ & 0.224 & 0.219 & 0.228\\
$\lambda_{1,3}$ & 0.102 & 0.099 & 0.105\\
  $\lambda_{1,4}$ & 0.451 & 0.445 & 0.456\\
$\lambda_{1,5}$ & 0.060 & 0.058 & 0.063\\
  $\lambda_{1,6}$ & 0.099 & 0.096 & 0.103\\
  \hline
$\lambda_{2,1}$ & 0.395 & 0.386 & 0.404\\
  $\lambda_{2,2}$ & 0.208 & 0.201 & 0.216\\
$\lambda_{2,3}$ & 0.103 & 0.098 & 0.109\\
  $\lambda_{2,4}$ & 0.090 & 0.084 & 0.096\\
$\lambda_{2,5}$ & 0.139 & 0.133 & 0.145\\
  $\lambda_{2,6}$ & 0.064 & 0.060 & 0.069\\
  \hline
$\lambda_{3,1}$ & 0.198 & 0.177 & 0.219\\
  $\lambda_{3,2}$ & 0.278 & 0.254 & 0.303\\
$\lambda_{3,3}$ & 0.115 & 0.098 & 0.133\\
  $\lambda_{3,4}$ & 0.219 & 0.195 & 0.244\\
$\lambda_{3,5}$ & 0.090 & 0.074 & 0.106\\
  $\lambda_{3,6}$ & 0.099 & 0.083 & 0.115\\
\hline
\end{tabular}
\caption{\uvcet{} fitted parameters under the \spacetime{} model. $x_j$ and $y_j$ denote the spatial coordinates of source $j$, $w_j$ denotes the relative intensity. $\lambda_{j,k}$ denotes the relative intensity of source $j$ in time bin $k$. $j=1$ corresponds to \uvcet\,B, $j=2$ corresponds to \uvcet\,A and $j=3$ corresponds to the background.The "mean" column gives the posterior means of the corresponding parameters, and the "(q16,q84)" column gives the 16$\%$ and 84 $\%$ posterior quantiles.}
\label{tab:SpaceTimeUVCet}
\end{table}
\bsp	
\label{lastpage}
\end{document}